\documentclass[useAMS,usegraphicx,usenatbib]{mn2e}
\usepackage{times}

\title[The MGC bivariate brightness distribution]{The Millennium Galaxy
Catalogue: the space density and surface brightness distribution(s) of
galaxies}

\author[S.P.~Driver et al.]{S.P.~Driver,$^{1,2}$\thanks{E-mail:
spd@mso.anu.edu.au}\thanks{PPARC Visiting Fellow} J.~Liske,$^3$ N.J.G.~Cross,$^4$ R.~De~Propris$^{1,2}$
and P.D.~Allen$^1$ \\ $^1$Research School of Astronomy and
Astrophysics, Australian National University, Cotter Road, Weston, ACT
2611, Australia \\ 
$^2$Department of Physics, University of Bristol,
Tyndall Avenue, Bristol, BS8 1TL, UK \\
$^3$European Southern Observatory,
Karl-Schwarzschild-Str.~2, 85748 Garching, Germany \\ $^4$Department
of Physics and Astronomy, The Johns Hopkins University, 3400 North
Charles Street, Baltimore, MD 21218, USA}

\newcommand{\mpas}{~mag~arcsec$^{-2}$}
\newcommand{\spas}{mag~arcsec$^{-2}$}
\newcommand{\bmgc}{B_{\mbox{\tiny \sc MGC}}}

\begin{document}

\date{Accepted
...... Received .....}

\pagerange{\pageref{firstpage}--\pageref{lastpage}} \pubyear{2004}

\maketitle

\label{firstpage}

\begin{abstract}
We recover the joint and individual space density and surface
brightness distribution(s) of galaxies from the Millennium Galaxy
Catalogue. The MGC is a local survey spanning 30.9~deg$^{2}$ and
probing approximately one--two\mpas\ deeper than either the Two-Degree
Field Galaxy Redshift Survey (2dFGRS) or the Sloan Digital Sky Survey
(SDSS). The MGC contains $10\,095$ galaxies to $\bmgc < 20$~mag with
$96$ per cent spectroscopic completeness. For each galaxy we derive
individual $K$-corrections and seeing-corrected sizes. We implement a
joint luminosity-surface brightness step-wise maximum likelihood
method to recover the bivariate brightness distribution (BBD)
inclusive of most selection effects. Integrating the BBD over surface
brightness we recover the following Schechter function parameters:
$\phi^* = (0.0177 \pm 0.0015) h^3$~Mpc$^{-3}$, $M_{\bmgc}^* - 5 \log h
= (-19.60 \pm 0.04)$~mag and $\alpha =-1.13 \pm 0.02$. Compared to the
2dFGRS \citep{norberg02} we find a consistent $M^*$ value but a
slightly flatter faint-end slope and a higher normalisation, resulting
in a final luminosity density $j_{b_J} = (1.99 \pm 0.17) \times 10^8
\, h \, L_{\odot}$~Mpc$^{-3}$ --- marginally higher than, but
consistent with, the earlier 2dFGRS (\citealp{norberg02}), ESP
(\citealp{esp}), and SDSS $z=0.1$ \citep{blanton03} results. The MGC is
inconsistent with the SDSS $z=0$ result ($+3\sigma$) if one adopts
the derived SDSS evolution (\citealp{blanton03}).

The MGC surface brightness distribution is a well bounded
Gaussian at the $M^*$ point with $\phi^* = (3.5 \pm 0.1) \times
10^{-2} h^3$~Mpc$^{-3}$, $\mu^{e*} = (21.90 \pm 0.01)$\mpas\ and
$\sigma_{\ln R_e} = 0.35 \pm 0.01$. The characteristic surface
brightness for luminous systems is invariant to $M_{\bmgc} - 5 \log h
\approx -19$~mag faintwards of which it moves to lower surface
brightness. The surface brightness distribution also broadens
($\sigma_{\ln R_e} \approx 0.5 - 0.7$) towards lower luminosities. The
luminosity dependence of $\sigma_{\ln R_e}$ provides a new constraint
for both the theoretical development (Dalcanton, Spergel \& Summers
1997; Mo, Mao \& White 1998) and numerical simulations (e.g.,
\citealp{cole}) which typically predict a mass-independent
$\sigma_{\ln R_e} \approx 0.56 \pm 0.04$ (see \citealp{vitvitska} and
\citealp{bullock}).

Higher resolution (FWHM $\ll 1$~arcsec) and deeper ($\mu_{\mbox {\tiny
\sc lim}} \gg 26$\mpas\ in the $B$-band) observations of the local
universe are now essential to probe to lower luminosity and lower
surface brightness levels.
\end{abstract}

\begin{keywords}
galaxies: general --
galaxies: fundamental parameters --
galaxies: luminosity function, mass function --
galaxies: formation --
galaxies: evolution --
astronomical data bases: catalogues
\end{keywords}

\section{Introduction}
\label{introduction}
The recognised convention for representing the overall galaxy
population is the luminosity distribution or space density of galaxies
(see review by Binggeli, Sandage \& Tammann 1988 and more recently
\citealp{driver03}). This distribution is typically derived from a
magnitude limited redshift survey which is corrected for Malmquist
bias (e.g., Efstathiou, Ellis \& Peterson 1988) and fitted with the
three parameter Schechter function \citep{schechter}. Generally, the
Schechter function is found to be a formally good representation,
although some surveys (e.g., \citealp{esp}; \citealp{ssrs2}) have
hinted at an upturn at the faintest limits,\footnote{It is worth
noting that the luminosity distributions seen in rich clusters such as
A963, Coma and Virgo are also seen to turn up at a similar magnitude,
(see e.g.\ \citealp{driver94}; \citealp{trentham}; Driver, Couch \&
Phillipps 1998; \citealp{driver03}; and \citep{popesso}). Conversely
the local group luminosity function remains flat to the Jeans mass
limit (\citealp{karachentsev}).} where the statistics become poor
(typically $M_B - 5\log h \approx -16$~mag, see also
\citealp{blanton04}). The process of fitting a luminosity function
(LF) reduces the galaxy population to three crucial numbers: the
characteristic luminosity, $M^*$ (or $L^*$), the normalisation,
$\phi^*$, and the faint-end slope, $\alpha$.

Since the pioneering work of \cite{cfa} numerous measurements of these
three crucial parameters have been made (see \citealp{mgc1}). Among
the most recent are those derived from the Two-Degree Field Galaxy
Redshift Survey (2dFGRS; \citealp{norberg02}) and the Sloan Digital
Sky Survey (SDSS1\&2; \citealp{blanton01}; 2003a). However, although
individual surveys yield Schechter function parameters to high
accuracy, a comparison between independently published values shows
excessively large variations, indicative of strong systematic errors
(see \citealp{cross01}; \citealp{mgc1}; \citealp{lapparent}).
\cite{cross02} discuss the impact of surface brightness selection
effects on measurements of the LF (see also \citealp{spray};
\citealp{dalcanton}) and find that such effects {\em may} provide an
explanation for the variation seen. In \cite{mgc1} we compared a
number of recent $B$-band Schechter function values and used the
precision number counts of the Millennium Galaxy Catalogue -- with a
uniform isophotal detection limit of $26$\mpas\ in the $B$-band -- to
revise the normalisation parameters. This resulted in a reasonably
consistent global Schechter function with: $M^*_{b_J}-5\log h =
(-19.64 \pm 0.08)$~mag, $\phi^* = (1.64 \pm 0.005) \times 10^{-2}
h^3$~Mpc$^{-3}$ and $\alpha=-1.1 \pm 0.1$ (taken from
\citealp{driver03}).

Despite the apparent convergence brought about by \cite{mgc1} there
are two strong reasons to now go beyond the
monovariate Schechter function:

\noindent
(1) Numerical and semi-analytic simulations are sufficiently
developed that it is now possible to predict not just the luminosity
distribution but also the size distribution (see e.g.\ \citealp{mo};
Buchalter, Jimenez \& Kamionkowski 2001; \citealp{shen03}) and hence
the {\em bivariate} distribution of luminosity and size (or luminosity
and surface brightness, see e.g.\ \citealp{sodre93}; \citealp{boyce};
\citealp{dejong}; \citealp{cross02}). This bivariate brightness
distribution (BBD) can be fitted by a 6-parameter Cho\l oniewski
function \citep{chol} -- essentially a Schechter function with a
luminosity dependent Gaussian distribution in surface brightness (see
e.g.\ \citealp{dejong}; \citealp{cross02}) -- or compared directly
without confinement by any functional form. Either way, a multivariate
distribution provides a more stringent constraint than the monovariate
Schechter function (see also \citealp{driver04});

\noindent
(2) \cite{mgc1} essentially circumvent the surface brightness issues
raised by \cite{cross02} by allowing the normalisation of the LFs to
accommodate the missing flux from low surface brightness galaxies.
This approximation succeeds because the galaxies that dominate the
bright galaxy number counts are all around the $L^*$ value, where the
surface brightness distribution appears well defined (see
\citealp{driver99}, \citealp{cross01}). We also know from studies of
giant ellipticals (\citealp{kormendy}; \citealp{graham}), disk
galaxies (\citealp{dejong}), dwarf galaxies in the local group
(\citealp{mateo}), rich clusters (\citealp{ferguson};
\citealp{andreon}) and from early work on the global field BBD
(\citealp{driver99}; \citealp{cross01}; \citealp{blanton01};
\citealp{shen03}) that there exists a luminosity surface brightness
relation (see also \citealp{bell03} for the near-IR relation and a
discussion of its variation with wavelength). Together these imply
that the impact of surface brightness selection effects will be a
function of luminosity. Therefore recovering the full luminosity
distribution can only be done by simultaneously considering luminosity
{\em and} surface brightness -- a conclusion reached many times over
the past half-century (\citealp{zwicky}; \citealp{disney};
\citealp{impey97}).

An additional motivation is that given the size of contemporary
datasets, it becomes logical to start to explore multivariate
distributions to quantify trends such as the luminosity-surface
brightness relation \citep{driver99} or the colour-luminosity relation
(e.g., \citealp{baldry}; see also \citealp{blanton03b}).

Here we reconstruct the bivariate luminosity-surface brightness
distribution with careful attention to the impact of surface
brightness and size selection effects. In Section~\ref{data} we
describe the imaging and redshift data. In Section~\ref{cosmoke} we
describe the conversion of our catalogue from apparent to absolute
properties, the various selection boundaries and our methodology for
deriving the joint luminosity-surface brightness distribution. We
apply it to our data in Section~\ref{mgcbbd} and compare our results
to previous estimates of the BBD (\citealp{dejong};
\citealp{cross01}), the LF (\citealp{norberg02}; \citealp{esp};
\citealp{blanton01}; \citealp{blanton03}) and the surface brightness
or size distribution (\citealp{cross01}; \citealp{shen03}). We
summarise our key results in Section~\ref{conclusions}. Throughout we
assume $\Omega_{o}=0.3, \Omega_{\Lambda}=0.7$ as defined in
Section~\ref{cosmology}.

\section{Data}
\label{data}
\subsection{Photometric catalogue}
The Millennium Galaxy Catalogue (MGC) is a deep ($\mu_{\mbox{\tiny \sc
lim}} = 26$\mpas), wide-field ($30.88$~deg$^2$) $B$-band imaging survey
obtained with the Wide Field Camera on the 2.5-m Isaac Newton
Telescope. The survey region is a long, 35-arcmin wide strip along
the equator, covering from $9^{\rm h} 58^{\rm m} 28^{\rm s}$ to
$14^{\rm h} 46^{\rm m} 45^{\rm s}$ (J2000), and is fully contained
within the regions of both the 2dFGRS and SDSS Data Release 1 (DR1;
\citealp{abazajian03}).

\begin{figure*}


\centering\includegraphics[width=\textwidth]{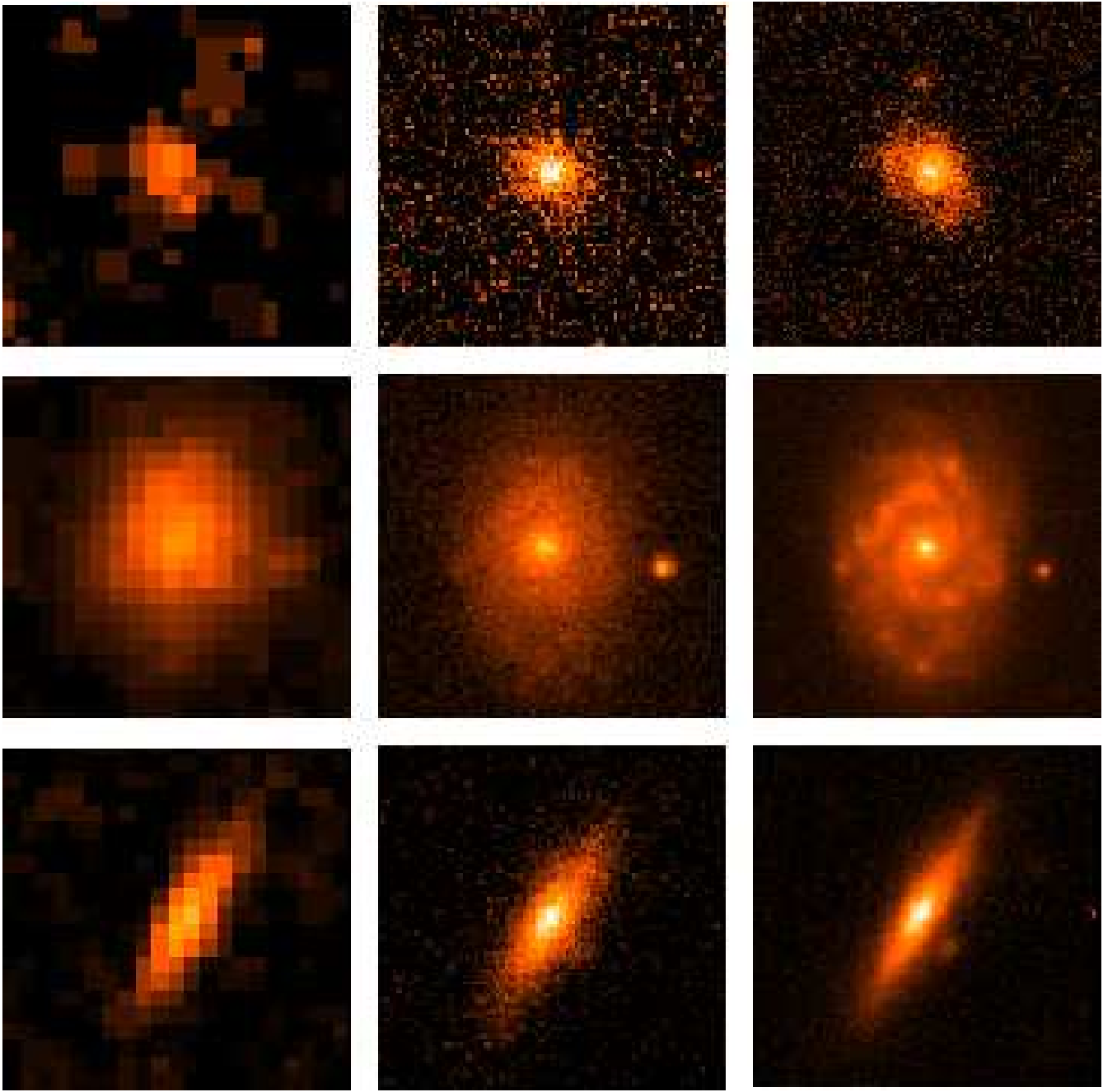}


\caption{A comparison between Digital Sky Survey (left panels, $b_{\rm
J}$-band), SDSS-DR1 (centre, $g$-band) and MGC (right panels) images:
MGC03494 (upper); MGC18325 (middle); MGC04795 (lower). The data
illustrate the greater depth and higher resolution of the MGC. All
images are $33''$ on each side and scaled to comparable dynamic
ranges.Figure degraded see http://www.eso.org/$\sim$jliske/mgc/ for full pdf
copy of this paper.}
\label{mgcexamples}
\end{figure*}

In \cite{mgc1} we gave a detailed description of the observations,
reduction, object detection, Galactic extinction correction and
classification (using SExtractor; \citealp{sex}) and presented
precision galaxy number counts over the range $16 \le \bmgc <
24$~mag. All non-stellar sources to $\bmgc = 20$~mag were visually
inspected and if an object was found to be incorrectly deblended or if
the object's parameters were obviously wrong, the object was
re-extracted by manually changing the SExtractor extraction parameters
until a satisfactory result was achieved. In addition, all low-quality
regions in the survey (e.g.\ near CCD defects) were carefully masked
out (see \citealp{mgc1} and \citealp{Lemon}). We demonstrated that the
internal photometric accuracy of the MGC is $\pm 0.023$~mag and that
the astrometric accuracy is $\pm 0.08$~arcsec. In \cite{mgc3} we then
compared the photometric accuracy, completeness and contamination of
the MGC to the 2dFGRS and SDSS-EDR and DR1 datasets. Being deeper and
of higher resolution, we found that the MGC is both more accurate and
more complete than either of these surveys. The MGC is currently the
highest quality and most complete representation of the nearby galaxy
population. Fig.~\ref{mgcexamples} shows images of three galaxies from
the Digital Sky Survey (left), the SDSS (centre) and the MGC (right),
illustrating the superior resolution and greater depth of the MGC data.

The MGC-BRIGHT catalogue of \cite{mgc1} was bounded at the bright end
by $16$~mag, where stars begin to flood in our best seeing fields. We
now extend this catalogue brightwards to $\bmgc = 13$~mag by carefully
examining all non-stellar objects in the range $13 \le \bmgc <
16$~mag, repairing objects that had been over-deblended by SExtractor
and checking completeness with NED at $\bmgc < 15$~mag, where
SExtractor -- optimised for faint galaxy detection and photometry --
may occasionally miss very large objects. A second change to
MGC-BRIGHT with respect to \cite{mgc1} is the deletion of some
exclusion regions. Liske et al.\ had placed exclusion regions around
very bright objects ($\bmgc < 12.5$~mag) to avoid spurious halo
detections and problems with background estimation in their
vicinity. $20$ such regions had been erroneously placed around
galaxies fainter than $12.5$~mag and these have been deleted,
resulting in an increase of the good-quality survey area from $30.84$
to $30.88$~deg$^2$.

Further minor changes were introduced to MGC-BRIGHT, each affecting
$\sim$tens of objects: different deblending decisions; spectroscopic
identification of galaxies previously misclassified as stars (compact
emission line galaxies) and vice versa (stars which appear extended
because of almost perfect alignment with small, faint background
galaxies); and identification of asteroids previously misclassified as
galaxies (discovered by comparison with SDSS-DR1 imaging data).

The final MGC-BRIGHT sample now comprises $10\,095$ galaxies. 

\subsection{Redshifts}
\subsubsection{Publicly available data}
By design the MGC survey region is fully contained within the 2dFGRS
and SDSS survey regions. Hence we first turned to these surveys in
order to obtain redshifts for MGC-BRIGHT. In addition we obtained
redshifts from the 2QZ \citep{2qz}, NED (excluding 2dFGRS and SDSS
objects) and the smaller surveys of Francis, Nelson \& Cutri (2004)
 (PF) and
\cite{impey96} (LSBG). In turn each survey, $X$, was matched against
MGC galaxies and stars with $\bmgc < 22$~mag. This magnitude limit and
the following matching procedure were chosen in order to be able to
account for every $X$ object and hence to verify the completeness of
the MGC.

\begin{figure}
\centering\includegraphics[angle=270,width=\columnwidth]{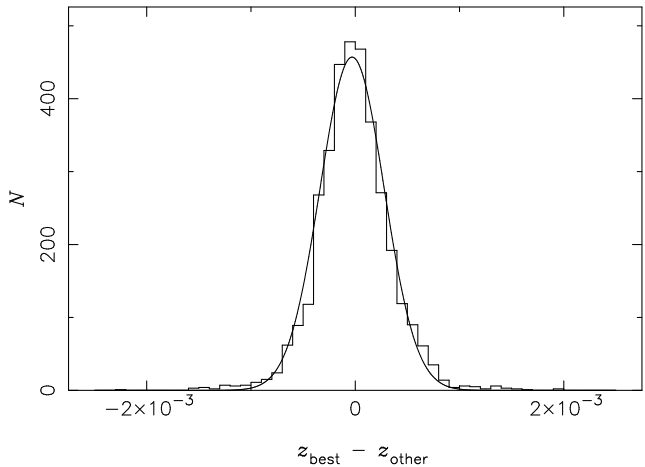}
\caption{Histogram of redshift differences for $2709$ objects with
multiple redshift measurements. The distribution has an rms dispersion
of $92$~km~s$^{-1}$. The solid line is a Gaussian with the same median
and rms as the data.}
\label{multiplezs}
\end{figure}

\begin{figure*}
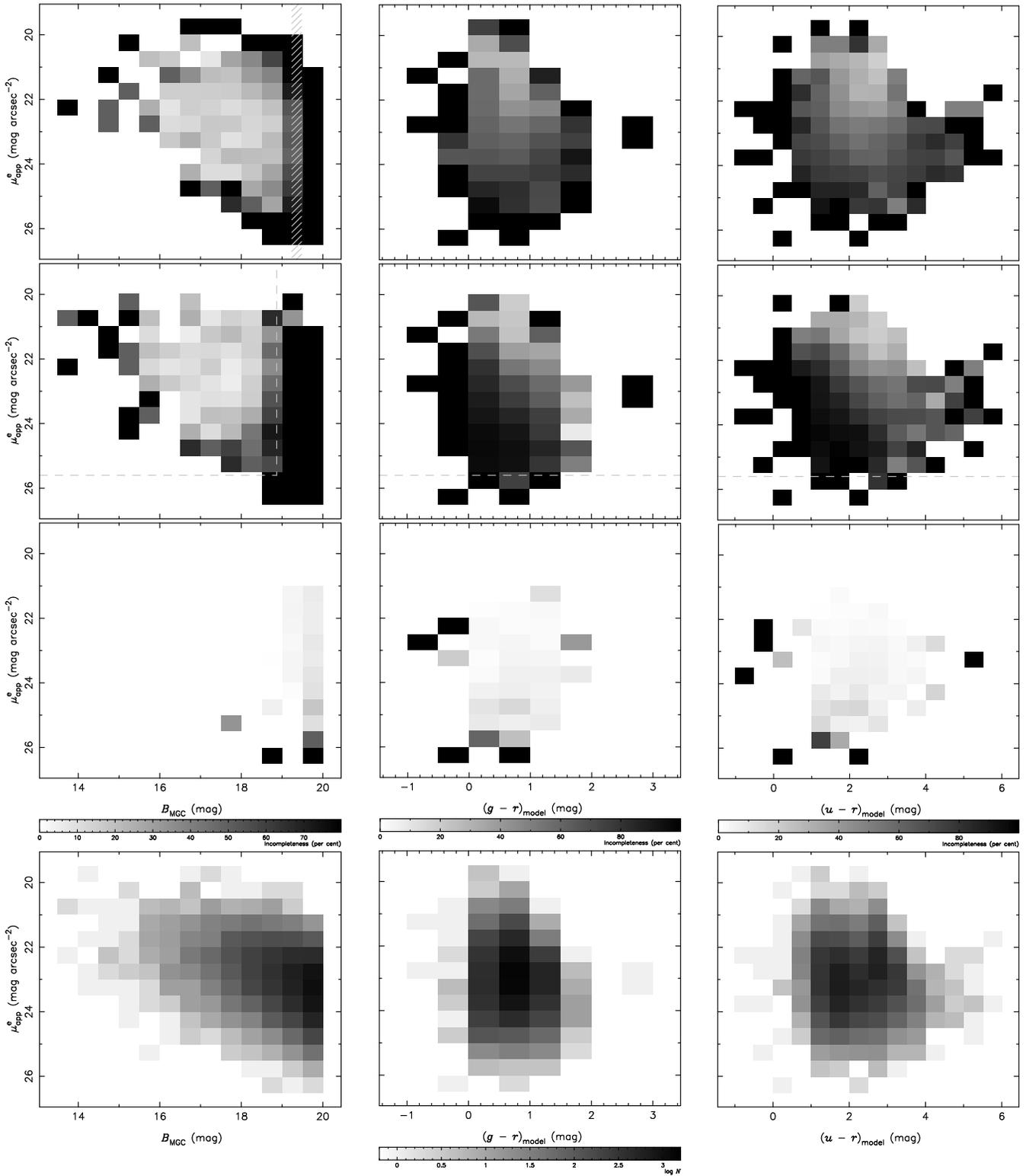

\hspace{-1.475\columnwidth}\includegraphics[angle=270,width=0.7\columnwidth]{2dfgrsinc_mu_b.ps}

\hspace{-1.475\columnwidth}\includegraphics[angle=270,width=0.7\columnwidth]{sdssinc_mu_b.ps}

\hspace{-1.475\columnwidth}\includegraphics[angle=270,width=0.7\columnwidth]{mgcinc_mu_b.ps}

\hspace{-1.475\columnwidth}\includegraphics[angle=270,width=0.7\columnwidth]{n_mu_b.ps}

\vspace{-20.0cm}

\includegraphics[angle=270,width=0.63\columnwidth]{2dfgrsinc_mu_g-r.ps}

\includegraphics[angle=270,width=0.63\columnwidth]{sdssinc_mu_g-r.ps}

\includegraphics[angle=270,width=0.63\columnwidth]{mgcinc_mu_g-r.ps}

\includegraphics[angle=270,width=0.63\columnwidth]{n_mu_g-r.ps}

\vspace{-20.5cm}

\hspace{1.4\columnwidth}\includegraphics[angle=270,width=0.63\columnwidth]{2dfgrsinc_mu_u-r.ps}

\hspace{1.4\columnwidth}\includegraphics[angle=270,width=0.63\columnwidth]{sdssinc_mu_u-r.ps}

\hspace{1.4\columnwidth}\includegraphics[angle=270,width=0.63\columnwidth]{mgcinc_mu_u-r.ps}

\hspace{1.4\columnwidth}\includegraphics[angle=270,width=0.63\columnwidth]{n_mu_u-r.ps}

\vspace{0.5cm}

\caption{Top row: the redshift incompleteness for the $10\,095$ galaxies in the
$30.88$~deg$^2$ MGC region using only 2dFGRS redshifts as a function of
apparent effective surface brightness and apparent magnitude (left),
($g-r$) (centre) and ($u-r$) colour (right). The second row shows the same
using only SDSS redshifts. These figures demonstrate the significant
contributions of these surveys to our final redshift catalogue (third row)
which has an overall completeness of $96.05$ per cent. The light hashed
region in the top left panel marks the approximate magnitude limit of the
2dFGRS, which varies as a function of position on the sky. The dashed
lines in the second row mark the approximate SDSS spectroscopic magnitude
and surface brightness limits. No galaxies beyond these respective limits
were targeted by these surveys. The bottom row shows the number of objects
per bin.}
\label{completeness}
\end{figure*}

First we found all pairs of $X$ and MGC objects where each is
identified as the other's nearest neighbour in that catalogue. A pair
with separation $d$ smaller than some $d_{\rm cut}$ is classified as a
match unless one or more other MGC objects have also identified the
$X$ object as their nearest neighbour in $X$ with a separation $< 2.5
\, d$ (where $d$ refers to the separation of the closest pair). These
$X$ objects as well as objects with $d \ge d_{\rm cut}$ and those
which have not been identified as a nearest neighbour by any MGC
objects are classified as initially unmatched. The value of $d_{\rm
cut}$ is chosen as the separation where the invariably occurring tail
in the distribution of separations begins to dominate.

The above definition of initially unmatched objects is designed to
conservatively flag all possible mismatches. Almost all mismatches
fall into one of three categories, which are a priori difficult to
distinguish from one another but for which one needs to proceed
differently: (i) unusually large positional disagreement; (ii) two (or
more) objects in one catalogue where the other catalogue only has one
object, because of different deblending choices or different
resolution; (iii) one object in $X$ but none in MGC because of
inaccurate photometry in either $X$ or MGC, or because the $X$ object
is an asteroid or a spurious noise detection, or because the object is
genuinely missing from the MGC.

All initially unmatched objects are visually inspected in order to
reliably identify those objects that were in fact correctly
matched. The few cases where the mismatch was due to poor object
extraction in the MGC were fixed.

Each redshift of a matched object is assigned a redshift quality,
$Q_z$, which has the same meaning as in the 2dFGRS \citep{2dfgrs}:
$Q_z = 1$ means no redshift could be measured; $Q_z = 2$ means
possible but doubtful redshift; $Q_z = 3$ means probable redshift; and
$Q_z = 4$ or $5$ means good or very good redshift. $Q_z = 1$ and $2$
are considered failures and we will use only redshifts with $Q_z \ge
3$ in this paper. In addition to the 2dFGRS this system is also
already used by PF. We translate the various quality parameters of the
other surveys to our system in the following manner: SDSS-DR1
redshifts are assigned $Q_z = 4$ unless the SDSS ZStatus or ZWarning
flags indicate a serious problem; for the 2QZ we use $Q_z = 5 - Q({\rm
2QZ})$; and NED and LSBG redshifts are assigned $Q_z = 3$.

In total the public data provided good quality redshifts for $4804$
MGC galaxies.

\subsubsection{MGCz}
In order to complete the redshifts for the full MGC-BRIGHT sample we
conducted our own redshift survey (which we label MGCz), targeting
those galaxies which did not already have publicly available
redshifts. These observation were mainly carried out with the Two
Degree Field (2dF) facility on the Anglo Australian Telescope. The
data were collected, reduced and redshifted in an identical manner to
that of the 2dFGRS (see \citealp{2dfgrs} for full details). In
particular we again used the same redshift quality parameter, $Q_z$,
as the 2dFGRS (see above).

The redshift incompleteness of the 2dF data showed a clear bias
against low-surface brightness galaxies. Hence we undertook additional
single-object long-slit observations, targeting low-surface brightness
galaxies ($\mu^e_{\rm app} > 24$\mpas, where $\mu^e_{\rm app}$ is the
apparent effective surface brightness within the half-light radius,
see Section~\ref{mgcsb}) as well as gaps in our 2dF coverage, with the
Double Beam Spectrograph (DBS) on the Australian National University's
$2.3$-m, the Low Resolution Spectrograph (LRS) on the $3.6$-m
Telescopio Nazionale Galileo on La Palma, the ESO Multi-Mode
Instrument (EMMI) on the New Technology Telescope in La Silla and the
Gemini Multiple Object Spectrograph (GMOS, long-slit nod \& shuffle
mode) on Gemini North. In all cases we observed between $600$ and
$3600$~s through a $1.5$ to $2$-arcsec wide slit, aligning the slit
with the major axis of the object to collect as much flux as
possible. All the data were reduced in a similar manner using standard
{\sc iraf} procedures. The resulting spectra are of similar resolution
as our 2dF data, but of higher quality: they increased our redshift
completeness at $\bmgc < 19.5$~mag and $\mu^e_{\rm app} > 24$\mpas\ from
$90.7$ to $98.8$ per cent.

\subsubsection{Combined redshift sample}
Using all available data and MGCz data qthe overall redshift
completeness of MGC-BRIGHT is $96.05$ per cent. In
Fig.~\ref{completeness} (third row from top) we show the final
incompleteness as a function of apparent effective surface brightness
and apparent magnitude, ($g-r$) and ($u-r$) model colours.  In
Table~\ref{alldata} we list the numbers of redshifts contributed by
each of the surveys.

For $2709$ galaxies we have more than one $Q_z \ge 3$ redshift
measurement, totaling $6226$ redshifts. These duplications are mainly
due to overlaps between the 2dFGRS and SDSS-DR1, but $232$ of these
galaxies have at least one MGCz redshift. For these objects we pick a
`best' redshift by sorting first by $Q_z$, then by signal-to-noise
ratio (where available) and then by the order of the surveys in
Table~\ref{alldata}. Fig.~\ref{multiplezs} shows the distribution of
differences between the multiple redshifts which has an rms of
$92$~km~s$^{-1}$.

\section{Cosmology, $K$-corrections, evolution, selection limits and
  analysis method}
\label{cosmoke}
To calculate the MGC BBD we first need to adopt a cosmological
framework, appropriate $K$-corrections [$K(z)$] and an evolutionary
model [$E(z)$]. It is of course possible to determine a LF while
simultaneously solving for one or more of these quantities -- see
e.g.\ \cite{blanton03}, who derive a LF while simultaneously solving
for both the number-density evolution and the luminosity
evolution. However, our concern with this approach is that it can lead
to erroneous results {\em if} the incompleteness is significant and
biased (as indicated in \citealp{mgc3} and Fig.~\ref{completeness}).
In this case the evolutionary functions would be fitting not only
evolution but also the uncorrected selection biases. Instead, our
philosophy is to adopt reasonable assumptions for the evolution and
concentrate on a strategy to account for selection bias.

\begin{figure}
\centering\includegraphics[width=\columnwidth]{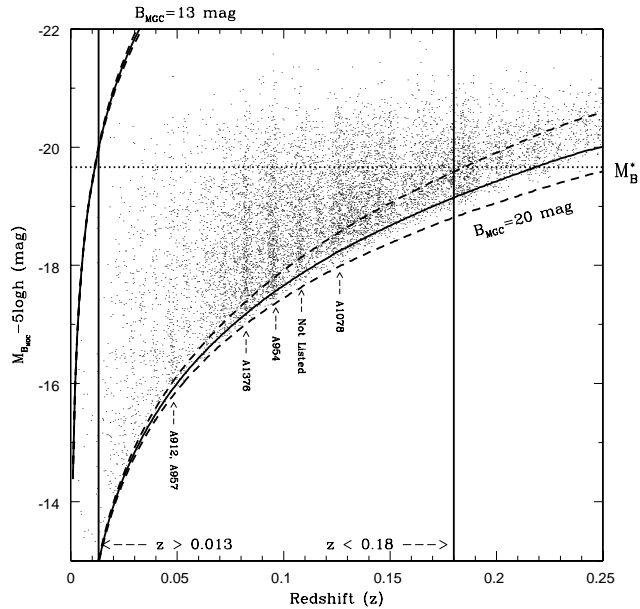}
\caption{Absolute magnitude versus redshift for the MGC using the
individual $K$-corrections as defined in
Section~\ref{kcorrections}. The solid curves show the bright and faint
magnitude limits assuming the mean $K$-correction and a universal
evolution of $L=L_{0} (1+z)^{0.75}$. The dashed lines again show the
magnitude limits but now assuming the bluest and reddest
$K$-corrections used in the MGC. The upper and lower redshift limits
are marked as vertical lines and the $M_B^*$ value from Norberg et
al.\ (2002) is shown as a horizontal dotted line. Note that the lower
redshift limit defines the minimum absolute magnitude of $M_B -
5 \log h = -13$~mag. The locations of known Abell clusters are
indicated and clearly show up as redshift over-densities.}
\label{Mz}
\end{figure}

\subsection{Cosmology}
\label{cosmology}
We use $\Omega_{0}=0.3$ and $\Omega_{\Lambda}=0.7$ -- broadly
confirmed by the WMAP mission combined with the 2dFGRS power spectrum
-- but adopt $h=H_0 / (100$~km~s$^{-1}$~Mpc$^{-1})$ for ease of
comparison with previous results. We also define maximum and minimum
redshift limits. The maximum limit is chosen as the point at which red
$L^*$-galaxies fall below our imposed $\bmgc = 20$~mag limit (this
occurs at $z_{\mbox{\tiny \sc max}}=0.18$, see Fig.~\ref{Mz},
excluding $1 774$ galaxies). The lower limit is determined by the
point at which the redshifts are significantly influenced by the local
velocity field. From \cite{jap} we find that the pairwise velocity
dispersion, as measured from the 2dFGRS, is $385$~km~s$^{-1}$. Hence
to ensure minimal impact from peculiar velocities (i.e.,
$v_{\mbox{\tiny Peculiar}} < 0.1 v_{\mbox{\tiny Recession}}$) we must
set $z_{\mbox{\tiny \sc min}}=0.013$ (excluding 44
galaxies). Fig.~\ref{Mz} shows the absolute magnitude versus redshift
distribution for our data along with the upper and lower redshift
limits (vertical lines) and the typical value of $M_B^*$
(\citealp{norberg02}; dotted line). Note that the lower $z$ limit,
combined with the faint apparent magnitude limit ($\bmgc = 20$~mag)
ultimately defines how far down the luminosity distribution one can
probe ($M_B - 5\log h \approx -13$~mag). One can only {\em reliably}
push fainter by extending the faint apparent magnitude limit of the
survey and not by covering a wider area (unless direct distances are
measured or the detailed local peculiar velocity field is modeled).
In total there are $7 878$ galaxies within these redshift limits.

\begin{figure}
\centering\includegraphics[width=\columnwidth]{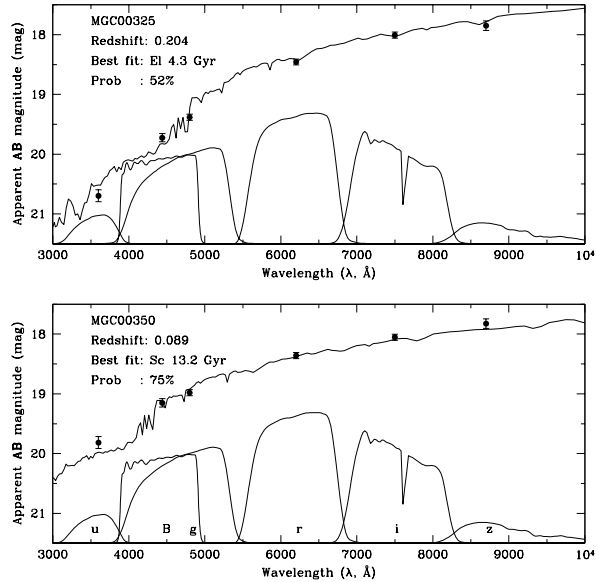}
\caption{Two examples of our spectral fitting algorithm, indicating a
good match to a young elliptical (upper panel) and older late-type
(lower). Once the optimum spectrum has been determined this is used to
determine the $\bmgc$-band $K$-correction.}
\label{kcorrexamples}
\end{figure}

\subsection{The $K$-correction}
\label{kcorrections}
$K$-corrections, or bandpass corrections, are galaxy specific ---
ultimately galaxy component specific --- and given the known variation
in galaxy types, a mean $K$-correction (as adopted in
\citealp{norberg02} and in most previous studies) may be overly simplistic
(see also \citealp{blanton03} who reach a similar conclusion). Here we
identify a best fit synthetic spectrum from the combined
broad-band MGC and SDSS-DR1 colours for each individual galaxy and
then directly measure the $K$-correction from this optimal
template. We use the extinction-corrected MGC Kron ($2.5$ Kron radii)
and extinction-corrected SDSS-DR1 Petrosian magnitudes for this
purpose. The Petrosian magnitudes are corrected to the AB system using
$u_{\mbox{\tiny \sc AB}}=u_{\mbox{\tiny \sc SDSS}}-0.04$;
$g,r,i_{\mbox{\tiny \sc AB}}=g,r,i_{\mbox{\tiny \sc SDSS}}$ and
$z_{\mbox{\tiny \sc AB}}=z_{\mbox{\tiny \sc SDSS}}+0.02$ as
recommended on the SDSS-DR2 website. The $\bmgc$ magnitudes are
adjusted for the known zeropoint offset of $0.039$~mag between the MGC
and SDSS-DR1 \citep{mgc3} and converted from Vega to AB. This
conversion was found to be $\bmgc(\mbox{AB}) = \bmgc(\mbox{Vega}) -
0.118$ for the MGC system.

\begin{figure}
\centering\includegraphics[width=\columnwidth]{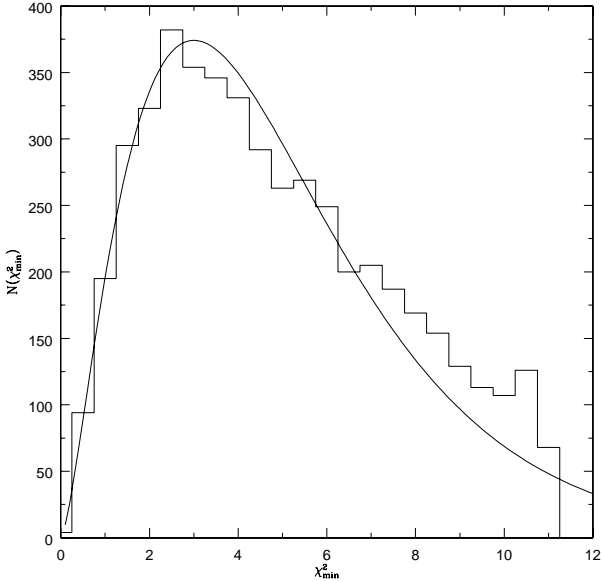}
\caption{The resulting $\chi^2_{\mbox{\tiny min}}$-probability
distribution from our $K(z)$ fitting algorithm (histogram) as compared
to the expected distribution (for five degrees of freedom, solid
line). Note that because we re-analyse galaxies with probabilities
less than $5$ per cent no values with $\chi^2_{\mbox{\tiny min}} \geq
11$ occur.}
\label{prob}
\end{figure}

We now wish to derive the optimal spectral template match to these
six filters in the AB system. We elect to use the spectrum library of
\cite{poggianti97} which includes ellipticals and early and late-type
spirals, with ages ranging from $15$~Gyr to $3.7$~Gyr ($27$ synthetic
templates in all). To determine the best template we calculate the
AB magnitudes as if observed through the $\bmgc$ and SDSS $u,g,r,i,z$
filters for a single template and iteratively rescale the amplitude of the
spectral template to minimise $\chi^2$:
\begin{center}
$\chi^2 = \sum_{x=u,B,g,r,i,z}[\frac{(x-x_{\mbox{\tiny \sc
model}})}{\Delta x}]^2$
\end{center}
Where the errors are given by: $\Delta u=\pm 0.10; \Delta B=\pm 0.07;
\Delta g=\pm 0.05; \Delta r=\pm 0.05; \Delta i=\pm 0.05$ and $\Delta
z=\pm 0.08$~mag.  These error values were derived as follows. Firstly,
using just the SDSS filters, we determined the $\chi^2_{\mbox{\tiny
min}}$-distribution for the optimal spectral fits based on the errors
listed on the SDSS website ($\Delta u =\pm 0.1$~[red leak], $\Delta gri
=\pm 0.02, \Delta z =\pm 0.03$). We then scaled the $griz$ errors until
the recovered $\chi^2_{\mbox{\tiny min}}$-distribution matched the
expected distribution for four degrees of freedom. We then repeated
this process now including the $\bmgc$ filter (see
Fig.~\ref{prob}). These final errors are in good agreement with the
study of \cite{mgc3}.

%
%
%

The $\chi^2$-minimisation was repeated for each of the $27$ spectral
templates and the smallest of the $27$ $\chi^2_{\mbox{\tiny min}}$
values used to identify the most appropriate template. Finally if the
$\chi^2_{\mbox{\tiny min}}$-probability for the optimal template was
less than $5$ per cent we repeated the process, first rejecting the
$\bmgc$ filter and then the $u$ and $z$ filters in turn (hence the
truncation at $\chi^2_{\mbox{\tiny min}} \approx 11$ on
Fig.~\ref{prob}). Magnitudes fainter than the quoted $95$ per cent
completeness limit were not used ($u=22.0$, $g=22.2$, $r=22.2$,
$i=21.3$, $z=20.5$~mag, see
\citealp{abazajian03}). Fig.~\ref{kcorrexamples} shows two example
galaxies with the best-fit spectral templates overlaid on the measured
magnitudes. Fig.~\ref{prob} shows the resulting $\chi^2_{\mbox{\tiny
min}}$ distribution (histogram) and the expected $\chi^2$-distribution
(for five degrees of freedom, solid line).

\begin{figure}
\centering\includegraphics[width=\columnwidth]{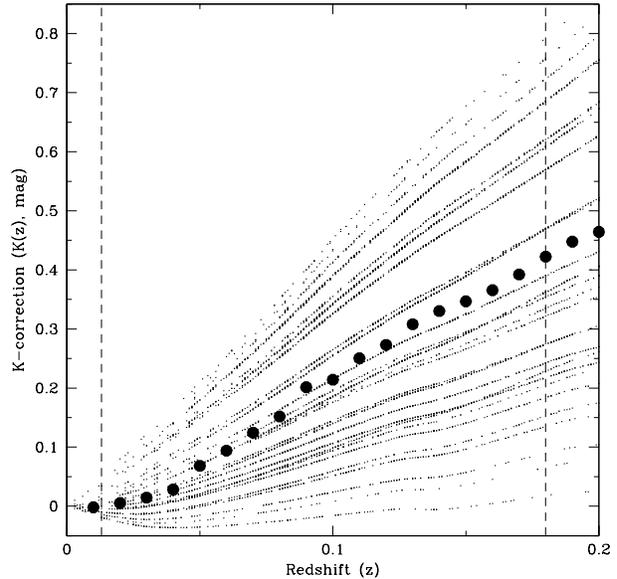}
\caption{Our final $K$-corrections. The MGC $K$-corrections trace out
the 27 tracks of the synthetic spectra of Poggianti (1997). The filled
circles indicate the mean of our data and the vertical dashed lines
show the adopted redshift limits used in the BBD analysis.}
\label{kcorr}
\end{figure}

Shown in Fig.~\ref{kcorr} are the resulting $K$-corrections for the
MGC, which trace out the $27$ different spectral templates, as well as
the mean $K$-correction in $0.01$ redshift intervals (large dots).
Finally, in the few cases ($25$ galaxies in total) where a galaxy did
not have an SDSS-DR1 match we assigned the mean $K$-correction value
at the galaxy's redshift. 

\subsection{Evolution}
\label{evolution}
We adopt a global evolution of the form $L = L_{0} (1+z)^{\beta}$ with
an initial value of $\beta=0.75$ (i.e., $E(z) = -0.75 \times
2.5\log(1+z)$, see \citealp{phillipps95}).  Given the redshift limits
this implies an evolutionary correction in the range $0.01 -
0.14$~mag. This is a passive luminosity evolution model with no
merging over this redshift range ($0.013 < z < 0.18$). In
Section~\ref{evolmstar} we solve for $\beta$ and find $-2.0 < \beta <
1.25$ (see Fig.\ \ref{beta}). More complex evolutionary scenarioes
will be considered in future papers where we sub-divide the galaxy
population by type and component. However we do note that recent
studies (e.g., \citealp{patton}) including our own (\citealp{mgc6})
indicate a very low low-z merger rate based on counts of dynamically
close pairs.  It is also worth noting that the step-wise maximum
likelihood method used assumes no number density evolution.

\subsection{Estimating MGC effective surface brightnesses}
\label{mgcsb}
We adopt the effective surface brightness, i.e.\ the average surface
brightness within the half-light radius, as our structural
measurement. The effective surface brightness is chosen as it makes no
assumption of the radial flux profile and can be measured empirically
from the data for any galaxy shape. Hence it is a direct measure
reflecting the global compactness regardless of whether the galaxy is
smooth or lumpy.

The half-light radius is the semi-major axis of the ellipse containing
half the flux of SExtractor's {\sc best} magnitude. The ellipse's
centre, ellipticity and position angle are taken from SExtractor. For
compact objects the half-light radius is affected by the
seeing. Through simulations of both exponential and de Vaucouleurs
profile galaxies using {\sc iraf artdata} (see Appendix~\ref{appsim}
for details) we find that we can correct the observed half-light
radius for this blurring and recover its true value, $r^0_e$, to an
accuracy of $10$ per cent (see Fig.~\ref{selection}) by:
\begin{equation}
r^0_e=\sqrt{r^2_e - 0.32 \, \Gamma^2},
\end{equation}
where $r_e$ is the observed half-light radius and $\Gamma$ is the FWHM
of the seeing profile, all in arcsec.


\begin{figure*}
\centering\includegraphics[width=\textwidth]{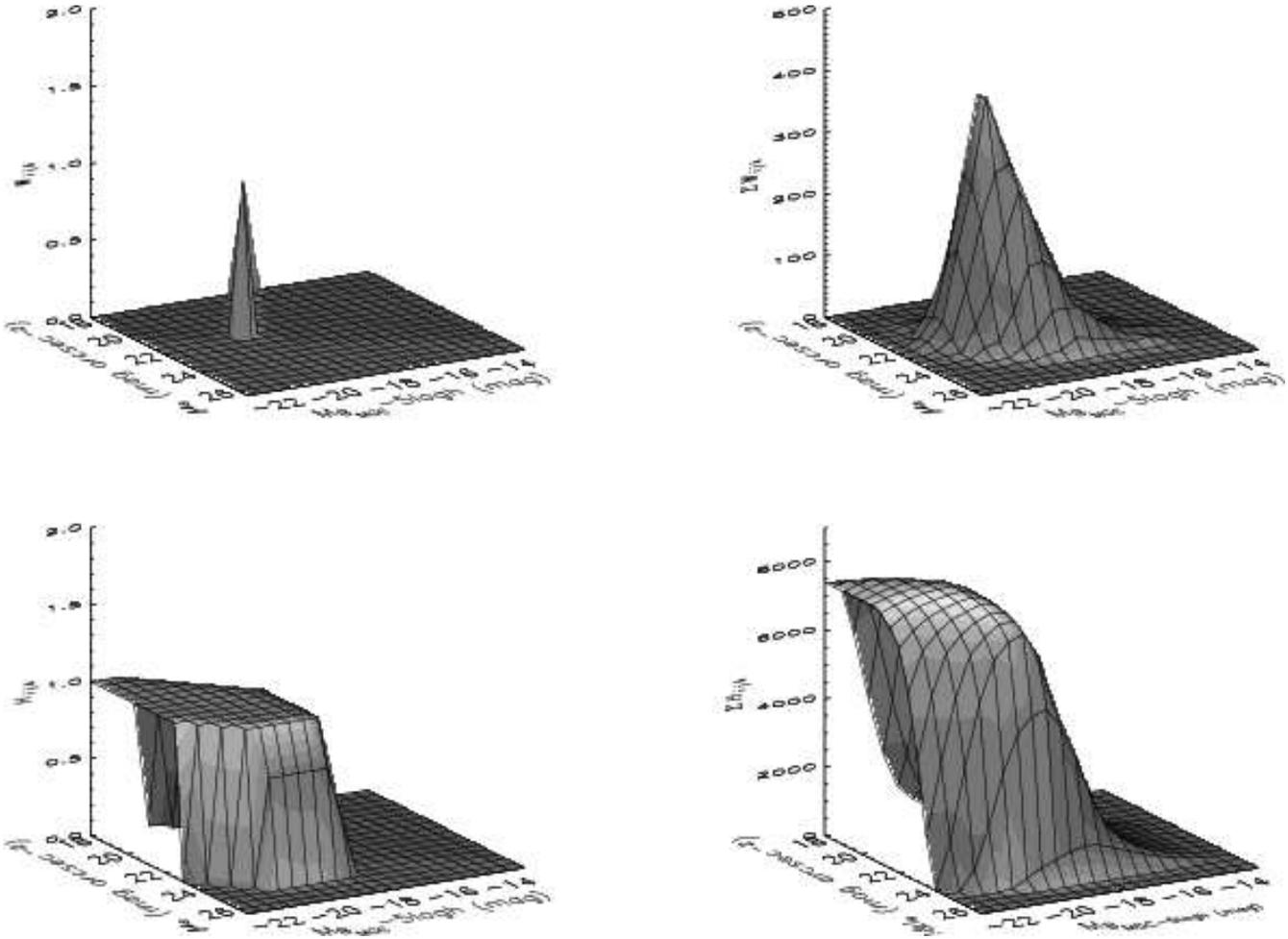}
\caption{An illustration of our BBD SWMl approach. Left panels:
examples of the individual matrices $W_{ijk}$ (weighting function;
upper) and $H_{ijk}$ (pseudo-visibility function; lower) for
MGC16504. Right panels: $\sum_i W_{ijk}$ (upper) and $\sum_i H_{ijk}$
(lower).Figure degraded see http://www.eso.org/$\sim$jliske/mgc/ for full
pdf copy of this paper.}

\label{matrices}
\end{figure*}

Work is currently underway to determine 2D bulge-disk decompositions
from the MGC data which will be presented in a future paper. However
at this stage it is unclear whether the effective radius or
independent bulge and disk parameters are the more fundamental. Here,
we define our absolute effective surface brightness, $\mu^e$,
as:
\begin{equation}
\mu^e = \bmgc + 2.5\log [2 \pi (r^0_e)^2] - 10 \log (1+z) - K(z) -
E(z),
\end{equation}
We note that $r^0_e$ is defined along the semi-major axis and hence
the above prescription assumes galaxies are optically thin, i.e.,
derived parameters are independent of inclination. The role
of dust will be explored in a future paper.

\subsection{Selection limits}
\label{sellim}
Like all surveys our catalogue will suffer from some level of
incompleteness in both the imaging and spectroscopic surveys. As our
data are the deepest available over this region of sky we cannot
empirically quantify our imaging incompleteness (see \citealp{mgc3}
for an assessment of the incompleteness of the SuperCOSMOS Sky Survey,
2dFGRS and SDSS-EDR/DR1 relative to the MGC). We can, however, define
the selection limits at any redshift, following \cite{driver99}, and
use this information later (Section~\ref{swml}) to quantify the
selection boundary in our final space density distribution.

\begin{figure}
\centering\includegraphics[width=\columnwidth]{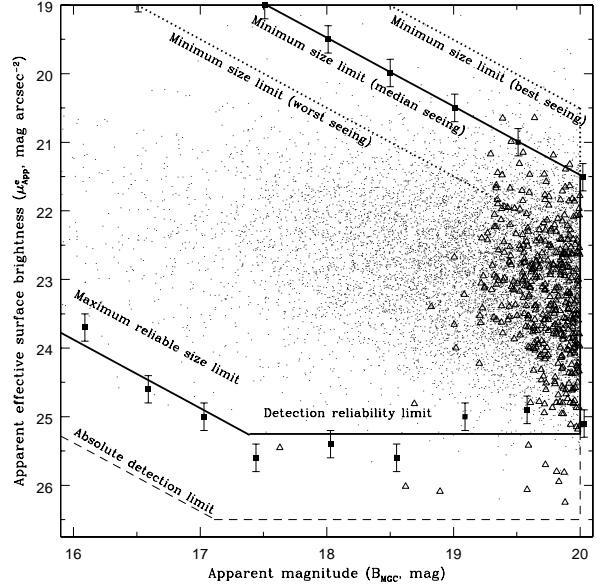}
\caption{The distribution of galaxies with (dots) and without (open
triangles) redshifts along with selection limits. Square data points
with errorbars show the selection limits as determined from the
simulations of Appendix~\ref{appsim} and the solid line shows the fit.
The low surface brightness and maximum size limits are defined as
limits of reliable photometry by requiring errors in the photometry of
less than $\pm 0.1$~mag and $\pm 0.1$\mpas. The dashed line is the
fundamental detection limit.}
\label{maxhlr}
\end{figure}

We begin by noting the MGC's apparent survey limits: a bright and
faint apparent magnitude limit of $m_{\mbox{\tiny \sc bright}}=13$~mag
and $m_{\mbox{\tiny \sc faint}}=20$~mag, a limiting apparent effective
surface brightness of $\mu^e_{\mbox{\tiny \sc lim}}=25.25$\mpas, a
minimum size of $r_{\mbox{\tiny \sc min}}= (0.04 \Gamma^2 + 0.37
\Gamma + 0.1)^{\frac{1}{2}}$~arcsec (where $\Gamma$ is the seeing
FWHM) and a maximum size of $r_{\mbox{\tiny \sc max}}= 15$~arcsec. The
last three limits refer to the seeing-corrected quantities (see
Section~\ref{mgcsb}) and were derived from simulations of
$45^\circ$-inclined optically thin exponential disks using {\sc
iraf}'s {\sc artdata} package (see Appendix~\ref{appsim} for
details). It is important to note that the surface brightness and
maximum size limits do {\em not} define the {\em absolute detection}
limits but the boundary of {\em reliable photometry} (where `reliable'
is arbitrarily defined as $\pm 0.1$~mag and $\pm 0.1$\mpas\ accuracy,
see Fig.~\ref{selection}).  Fig.~\ref{maxhlr} shows these limits in
the apparent magnitude-apparent effective surface brightness plane,
along with the galaxy distribution with (dots) and without (triangles)
redshifts.


\begin{figure}
\centering\includegraphics[width=\columnwidth]{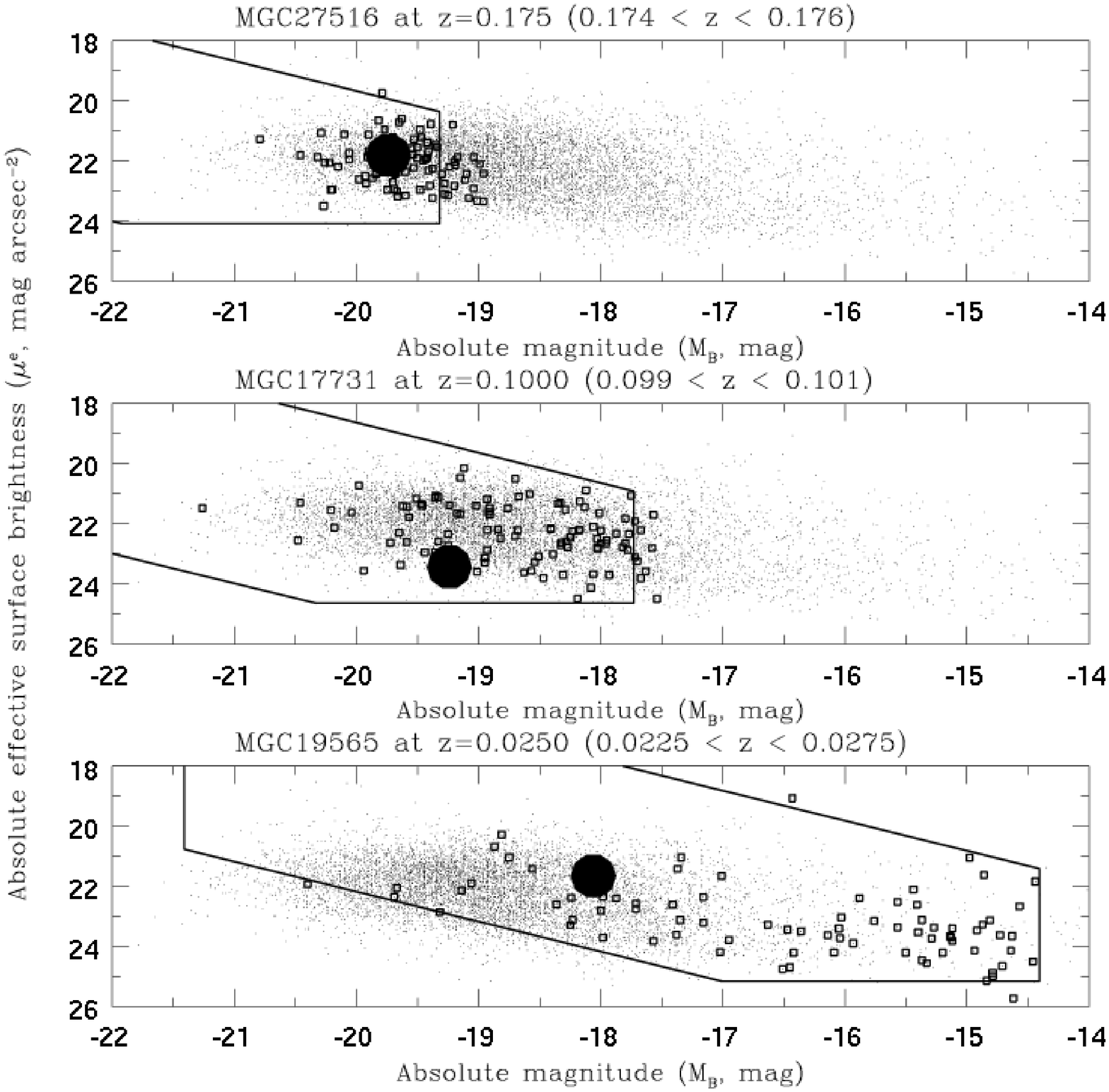}
\caption{The selection boundaries for three galaxies as indicated
along with the location of the galaxy (solid circle), those with
similar redshifts (open squares but differing $K$-corrections) and the
full ($0.013 < z < 0.18$) MGC distribution (dots). The figure
illustrates that the selection boundaries vary dramatically with
redshift (bottom to top) such that each redshift slice, within the
survey limitations, only samples a restricted region of the
luminosity-surface brightness plane. Note that some points may lie
outside the box if their $K$-corrections are significantly different
from the galaxy for which the selection lines have been drawn.
Figure degraded see http://www.eso.org/$\sim$jliske/mgc/ for full pdf
copy of this paper.}
\label{limfig}
\end{figure}

For each galaxy $i$ with known redshift, $z_i$, we can now determine
the parameter space in absolute magnitude, $M$, and absolute effective
surface brightness, $\mu^e$, over which this galaxy could have been
observed, given the survey limits. The appropriate limits are
(following \citealp{driver99}):
\begin{eqnarray}
M_{{\mbox{\tiny \sc bright},i}} & = & m_{\mbox{\tiny \sc bright}} - 5
\log d_L(z_i) - 25 - K(z_i) - E(z_i)\nonumber\\
M_{\mbox{\tiny \sc faint},i} & = & m_{\mbox{\tiny \sc faint}}-5 \log
d_L(z_i) - 25 - K(z_i) - E(z_i) \nonumber\\
\mu^e_{\mbox{\tiny \sc low},i} & = & \mu^e_{\mbox{\tiny \sc lim}} - 10
\log (1+z_i) - K(z_i) - E(z_i) \nonumber\\
\mu^e_{\mbox{\tiny \sc low},i} & = & M_i + 5
\log d_L(z_i) + 25 + 2.5\log (2 \pi r_{\mbox{\tiny \sc max}}^2)\\
& & \mbox{} - 10 \log (1+z_i) \nonumber\\
\mu^e_{\mbox{\tiny \sc high},i} & = & M_i + 5
\log d_L(z_i) + 25 + 2.5\log (2 \pi r_{\mbox{\tiny \sc min}}^2)\nonumber\\
& & \mbox{} - 10\log (1+z_i) \nonumber
\end{eqnarray}
Note that $\mu^e_{\mbox{\tiny \sc low},i}$ is defined by both the
minimum effective surface brightness for reliable detection,
$\mu^e_{\mbox{\tiny \sc lim}}$, and the maximum reliable size limit,
$r_{\mbox{\tiny \sc max}}$. The maximum reliable size limit is imposed
by the smoothing box size when calculating the background sky
levels. These limits are introduced in more detail in
\cite{driver99}. One can also envisage a sixth limit (also
$\mu^e_{\mbox{\tiny \sc high},i}$) due to the dynamic range of the CCD
detector. In our case none of the galaxies have their central regions
flooded and so this limit can be ignored.

Fig.~\ref{limfig} shows the selection limits for three galaxies at
redshifts $0.025$ (lower panel), $0.100$ (middle) and $0.175$ (upper)
along with the location of the galaxy (large solid circle), galaxies
with similar redshifts (open squares) and the full ($0.013 < z <
0.18$) MGC distribution (dots). The selection boundary consistently
forms a five-sided figure which glides through the full distribution
as the redshift increases. Because of the individual $K$-corrections
each galaxy has its own unique selection window. It is worth noting
that large luminous galaxies are undetectable, or at least
photometrically unreliable, at very low redshift. Hence the methods of
\cite{shen03} which use $z_{\mbox{\tiny \sc median}}(M,\mu^e)$-style
corrections, {\em may} underestimate this population because they
implicitly assume selection boundaries only cut into the distribution
at some upper redshift limit. \cite{cross01} accounted for this effect
in their analysis.

For each galaxy with known redshift we now define an observable window
function, $O_i(M,\mu^e)$, as:
\begin{equation}
O_i(M, \mu^e) = \left\lbrace
\begin{array}{lcl}
1 & \mbox{if} & M_{\mbox{\tiny \sc bright},i} < M < M_{\mbox{\tiny \sc
    faint},i} \\
& \mbox{and} & \mu^e_{\mbox{\tiny \sc high},i} < \mu^e <
           \mu^e_{\mbox{\tiny \sc low},i}\\
\\
0 & \multicolumn{2}{l}{\mbox{otherwise.}}
\end{array} \right.
\end{equation}
The functions $O_i$ will be used in Section~\ref{swml} in the
construction of the BBD.

\subsection{Bivariate step-wise maximum likelihood}
\label{swml}
We now construct the MGC BBD using a bivariate brightness step-wise
maximum likelihood (SWML) estimator similar to that described by
\cite{sodre93}. This is an extension of the standard SWML method
defined by \cite{eep} in the sense that the sample is now divided into
bins of both absolute magnitude {\em and} absolute surface brightness
as opposed to just absolute magnitude alone. Note that \cite{sodre93}
formulated their method in terms of absolute magnitude and a physical
diameter. However, the MGC is primarily magnitude and surface
brightness limited with a fixed isophotal detection limit of
$26$\mpas. For this reason we elect to work in $M$-$\mu^e$ space
rather than $M$-$R_e$ space.

Essentially, for $i=1...N$ objects the volume-corrected relative BBD in
$j=1...J$ absolute magnitude bins and $k=1...K$ surface brightness
bins with widths $\Delta M$ and $\Delta \mu^e$ can be evaluated by
(cf.\ \citealp{sodre93}):
\begin{equation}
\label{phijk}
\phi_{jk}=\frac{\sum_i^N W_{ijk}} 
{\sum^N_i [H_{ijk}/\sum^L_l\sum^M_m \bar \phi_{lm} H_{ilm}]},
\end{equation}
where the relative space density of galaxies, $\phi_{jk}$, is given in
terms of the space density of the previous iteration, $\bar
\phi_{jk}$, the weighting function, $W_{ijk}$, and the visibility
function, $H_{ijk}$.

The weighting function, $W_{ijk}$, corrects for the incompleteness in
our redshift survey. This incompleteness is demonstrably not random
(see Fig.~\ref{completeness}) and we are preferentially missing faint
dim galaxies with extreme colours. When constructing LFs this bias is
often ignored (e.g., \citealp{blanton03}) or considered a function of
magnitude only (e.g., \citealp{norberg02}).  Here, we account for
luminosity and surface brightness dependent incompleteness and define
$W_{ijk}$ as:
\begin{equation}
W_{ijk} = \left\lbrace
\begin{array}{lcl}
\frac{N_i}{N_i(Q_z \ge 3)} & \mbox{if} & \! \! M_j-{\frac{\Delta M}{2}} 
\leq M_i < M_j+{\frac{\Delta M}{2}}\\
& \mbox{and}& \mu^e_k-\frac{\Delta \mu^e}{2} \leq \mu^e_i < \mu^e_k +
\frac{\Delta \mu^e}{2}\\
\\
0 & \multicolumn{2}{l}{\mbox{otherwise,}}
\end{array} \right.
\end{equation}
where $N_i$ is the total number of galaxies lying in the same apparent
magnitude-apparent surface brightness bin as galaxy $i$ and $N_i(Q_z
\ge 3)$ is the number of galaxies with known redshifts (i.e., those
with $Q_z \ge 3$, see Section~\ref{data}) in the same bin. Hence each
galaxy is weighted by the inverse of the redshift completeness in the
galaxy's apparent magnitude-surface brightness bin.

The visibility function $H_{ijk}$ is given by:
\begin{equation}
H_{ijk}=\frac{1}{(\Delta M \Delta \mu^e)}\int_{M_j-\Delta
  M/2}^{M_j+\Delta M/2} \!\!\!\!\! {\rm d}M \int_{\mu^e_k-\Delta
  \mu^e/2}^{\mu^e_k+\Delta \mu^e/2} \!\!\!\!\! {\rm d}\mu^e
  O_i(M,\mu^e),
\end{equation}
where $O_i(M, \mu^e)$ is defined in Section~\ref{sellim}. Hence
$H_{ijk}$ is the fraction of the $M_j$-$\mu^e_k$ bin which lies inside
the observable window of galaxy $i$.

Fig.~\ref{matrices} shows example $W_{ijk}$ and $H_{ijk}$ matrices for
$i =$ MGC16504 (upper and lower left respectively) as well as the
summed $\sum_{i=1}^N W_{ijk}$ and $\sum_{i=1}^N H_{ijk}$ matrices
(upper right and lower right respectively). $\sum_i W_{ijk}$ is
essentially the observed galaxy distribution corrected for
incompleteness and $\sum_i H_{ijk}$ is analogous to the visibility
surface of Visibility Theory (Phillipps, Davies \& Disney 1990),
although not precisely as it does not account for the variation of
physical cross-section with redshift or for isophotal corrections.
The steep sides of the $\sum_i H_{ijk}$ surface (Fig.~\ref{matrices})
highlights the narrow range over which surface brightness selection
effects go from negligible to extreme and hence the importance of
understanding a survey's selection limits.

\section{The MGC bivariate brightness distribution}
\label{mgcbbd}
\begin{figure}
\hspace{-1.0cm}\includegraphics[height=8.0cm,width=10.0cm]{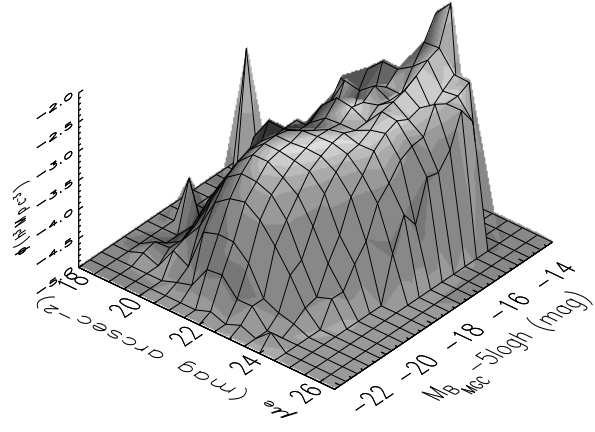}
\caption{The normalised bivariate brightness distribution. The
vertical axis is logarithmic in units of $h^3$~Mpc$^{-3}$~mag$^{-1}$
(\spas$)^{-1}$.}
\label{bbd}
\end{figure}

We now apply the above methodology to our MGC sample. To recap, we
extract galaxies in the range $13 < \bmgc < 20$~mag, determine
individual $K$-corrections, determine seeing-corrected surface
brightness estimates, attach an incompleteness weight to each galaxy,
reject galaxies which lie outside of their designated reliably
observable regions ($7$ galaxies too large, $99$ too compact and $22$
too dim), generate weighting and visibility matrices for each galaxy
and then iteratively apply equation~(\ref{phijk}) until a stable set
of $\phi(M, \mu^e)$-values is found. Fig.~\ref{bbd} shows the result
of this procedure after renormalisation (described below). It is worth
noting that the majority of the rejected galaxies are extremely
compact and have entered our sample only through spectroscopy of
apparently stellar objects (2QZ, SDSS QSO survey, NED and MGCz).

The SWML method, by its construction, loses all information regarding
the absolute normalisation (cf.\ \citealp{eep}). Here the
normalisation issues are compounded by the consideration of selection
bias. We overcome this by calculating the number of galaxies in the
absolute magnitude and absolute surface brightness interval $-20 < M_B
-5 \log h < -19.5$~mag and $20 < \mu^e < 24$\mpas. This parameter
space is observable for any $K(z)$ over the redshift range, $0.047 < z
< 0.162$ (equivalent to a co-moving volume of $311 \, 849 \,
h^{-3}$~Mpc$^3$ for the $30.883$~deg$^2$ field-of-view). Within these
redshift, absolute magnitude and surface brightness ranges there are
$785$ galaxies with a summed weight, due to incompleteness, of
$787.8$. The $\phi$ values are now rescaled to reproduce the correct
space density for these galaxies (i.e., $0.00253 \pm 0.00020 \,
h^3$~Mpc$^{-3}$).  The error includes both Possion and large scale
structure considerations and was derived from mock 2dF NGP catalogues
(\citealp{Cole98}; http://star-www.dur.ac.uk/$\sim$cole/mocks/main.html)
extracted from the Virgo Consortium Hubble Volume. Volumes equivalent
in shape to the MGC normalisation volume were extracted and the
standard deviation of $-20 < M_B -5 \log h < -19.5$~mag galaxies
determined. The final error, inclusive of large scale structure, is
roughly twice that of the Poisson error alone.

\begin{figure}
\hspace{-1.0cm}{\includegraphics[height=8.0cm,width=10.0cm]{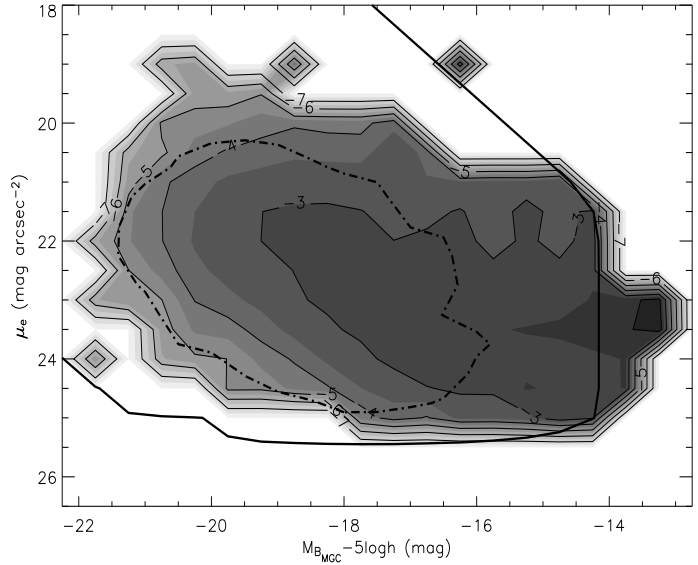}}
\caption{The final space density or bivariate brightness distribution
of galaxies shown as both a greyscale image and contours on
logarithmic scales in units of $h^3$~Mpc$^{-3}$~mag$^{-1}$
(\spas$)^{-1}$. The contour spacing is 1 dex in $\phi$ with the
greyscale at 0.5 dex intervals. The thick solid line denotes the
selection boundary defined as the BBD region where at least $100$
galaxies could have been detected and is equivalent to an effective
volume limit of $1800h^{-3}$Mpc$^{3}$. The thick dashed line encompasses
the region within which the statistical error is smaller than $25$ per
cent.}
\label{bbdcont}
\end{figure}

Fig.~\ref{bbdcont} shows the final normalised BBD on a logarithmic
scale. The thick dotted line shows the region within which the
statistical errors are $25$ per cent or less. The thick solid line
shows the effective detection limit defined as the BBD region sampled
by at least $100$ galaxies (equivalent to a volume limit of $\sim 1800
\, h^{-3}$~Mpc$^{3}$). The detection limit bounds the BBD at the low
surface brightness and faint limits and clearly starts to impinge on
the distribution from around $M_B - 5 \log h \approx -18$~mag
(coincidentally the usually adopted boundary between the dwarf and
giant galaxy populations). This confirms that even the MGC, the
deepest local imaging survey to date, remains incomplete for both high
($M_B - 5 \log h \approx -15$~mag) and low surface brightness
galaxies. To sample these regions requires both deeper
($\mu^e_{\mbox{\tiny \sc lim}} \gg 26$\mpas) {\em and} higher
resolution imaging data ($\Gamma \ll 1.27$~arcsec) over a comparable
or larger area ($\Omega \geq 30$~deg$^2$) with high spectroscopic
completeness.

Following \cite{dejong} and \cite{cross02} we attempted to fit the MGC
BBD with a Cho\l oniewski function \citep{chol}, given by:
\begin{eqnarray}
\lefteqn{ \phi(M,\mu^e) = \frac{0.4 \, \ln 10}{\sqrt{2 \pi} \,
    \sigma_{\mu^e}} \, \phi^* 10^{0.4 (M^*-M)(\alpha+1)}
  e^{-10^{0.4(M^*-M)}} } \nonumber\\
& & \times \exp\left\lbrace -\frac{1}{2}\left[\frac{\mu^e-\mu^{e*}-\beta(M-M^*)}{\sigma_{\mu^e}}\right]^2\right\rbrace
\end{eqnarray}
The parameters $\phi^*$, $M^*$ and $\alpha$ are the conventional
Schechter function parameters \citep{schechter}. The three additional
parameters, $\mu^{e*}$, $\sigma_{\mu^e}$ and $\beta$ are the
characteristic surface brightness, the Gaussian width of the surface
brightness distribution and the luminosity-surface brightness relation
respectively. This function has been derived analytically by
\cite{dss97}, \cite{mo} and \cite{dejong}. Note that the additional
normalisation factor, $(\sqrt{2 \pi} \sigma_{\mu^e})^{-1}$, ensures
that $\phi^*, M^*$ and $\alpha$ are directly equivalent to the more
familiar Schechter function parameters. The fit is achieved using the
downward simplex method \citep{nr} and the 1-$\sigma$ errors are
derived from a one-dimensional $\chi^2$-minimisation across each
parameter ($1\sigma \equiv \Delta \chi^2=158.9$). 

\begin{figure}
\hspace{-1.0cm}\includegraphics[height=8.0cm,width=10.0cm]{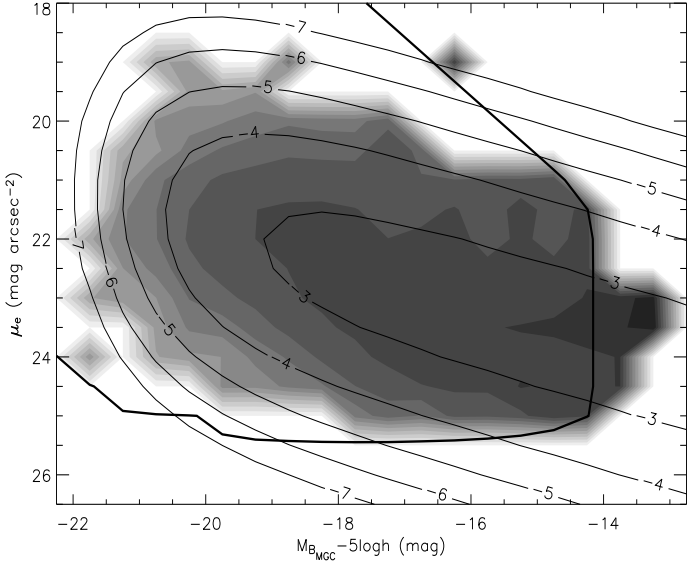}
\caption{The contours show the Cho\l oniewski function which best fits
our final BBD in units of $h^3$~Mpc$^{-3}$~mag$^{-1}$
(\spas$)^{-1}$. The greyscale image and the thick solid line are the
same as those in Fig.~\ref{bbdcont}.}
\label{bbdfit}
\end{figure}

For the $151$ degrees of freedom ($157$ data values minus $6$ model
parameters) we find an unreduced $\chi^2_{\mbox{\tiny min}}$ of
$608.7$. This suggests that the Cho\l oniewksi function is a poor fit
to the data (see Fig.~\ref{bbdfit}). The main reason appears to be the
intrinsic assumption of a constant $\sigma_{\mu^e}$, whereas the
distribution shown in Fig.~\ref{bbd} clearly broadens, a feature also
observed by \cite{shen03} in their SDSS data. We discuss this
further in Section~\ref{sbd}. Cho\l oniewski parameters and associated
errors are tabulated in the upper half of Table~\ref{lfdata} along
with other published data including our previous measurement based on
the 2dFGRS \citep{cross02}.

\subsection{The space density of galaxies}
\label{spacedensity}
\begin{figure}
\centering\includegraphics[width=\columnwidth]{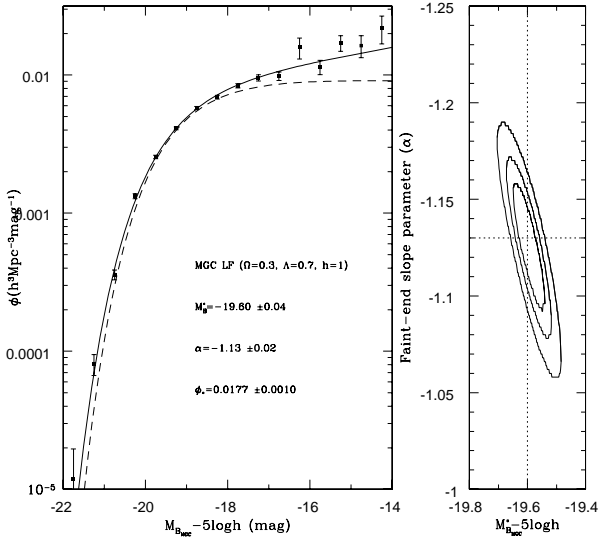} 
\caption{Left panel: the luminosity function derived by integrating
the BBD over surface brightness with the best-fit Schechter function
(solid line). Also shown is the implied Schechter function from the fitted
Cho\l oniewski function (dashed line). Right panel: the 1, 2 and
3-$\sigma$ contours of the Schechter function fit.}
\label{lf}
\end{figure}

Of particular interest over the past few decades has been the galaxy
luminosity distribution. This can be derived from the BBD by
integrating over the surface brightness axis, (the associated errors
are the root sum squares of the individual errors for each bin).
Fig.~\ref{lf} (left panel) shows the result: $\phi^* = (0.0177 \pm
0.0015) \, h^3$~Mpc$^{-3}$, $M_B^* - 5 \log h = (-19.60 \pm 0.04)$~mag
and $\alpha =-1.13 \pm 0.02$. Note that the error in the normalisation
parameter, $\phi^*$, is the combination (in quadrature) of the error
from the fitting algorithm combined with the error from the number of
galaxies within the normalisation volume.  For completeness we also
show the implied LF from our Cho\l oniewski function fit (dotted
line).  However recall the poor $\chi^2_{\mbox{\tiny min}}$ value from
this fit which highlighted the fit's inability to model the dwarf
population correctly (mainly due to the broadening of the surface
brightness distribution). The right panel of Fig.\ref{lf} shows the
formal 1,2,3-$\sigma$ error contours for the Schechter function fit to
the collapsed MGC BBD luminosity distribution.

\subsubsection{Comparison to the ESP, 2dFGRS and SDSS LFs}
\begin{figure}
\centering\includegraphics[width=\columnwidth]{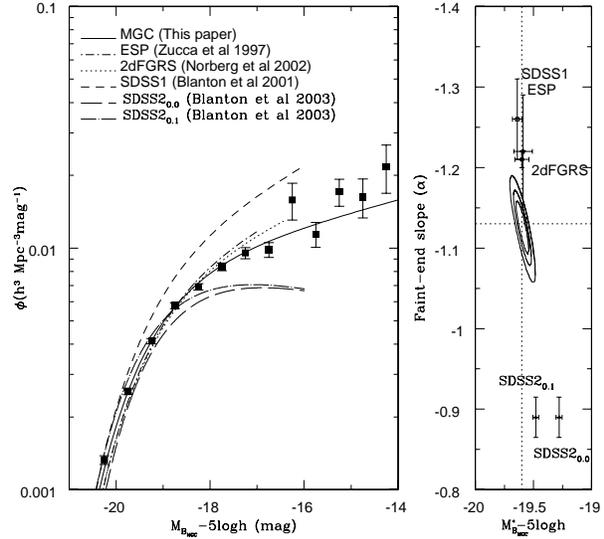}
\caption{A comparison of recent luminosity function estimates as
indicated (left panel) and the associated errors in $M^*$ and $\alpha$
together with the MGC $1$, $2$ and $3$-$\sigma$ contours (right
panel).}
\label{lfcomp}
\end{figure}

Fig.~\ref{lfcomp} compares our LF based on the BBD analysis of the
deeper, higher quality, higher completeness MGC data to a number of
other recent LF estimates (tabulated in the lower half of
Table~\ref{lfdata}). Note that these Schechter functions are shown
with their original $\phi^*$-values as opposed to the \cite{mgc1}
corrected values. We find our LF recovers a comparable $M^*$ value to
most previous surveys, but a slightly higher normalisation (as one
might expect from a BBD-style analysis which recovers more light from
the low surface brightness population than a non-BBD style analysis),
and an $\alpha$ value consistent with the median of these other
surveys. In comparison to the ESP LF (\citealp{esp}) we find
consistent Schechter function parameters within the quoted errors of
the two surveys. In comparison to the 2dFGRS LF (\citealp{norberg02})
we find consistent $M^*$ and $\phi^*$ values but a flatter (move
positive) $\alpha$. This discrepancy in $\alpha$ is formally
significant at the $3\sigma$-level, however we do note that the 2dFGRS
has undergone a major overhaul of its photometry since this result
(see discussion in \citealp{mgc3}). The final 2dFGRS
post-recalibration LF has yet to be determined. Our result are
consistent with the $M^*$ value only for SDSS1 and inconsistent with
all three Schechter function values for SDSS2.  A detailed discussion
of this discrepancy follows in the next section.

\begin{figure}
\centering\includegraphics[width=\columnwidth]{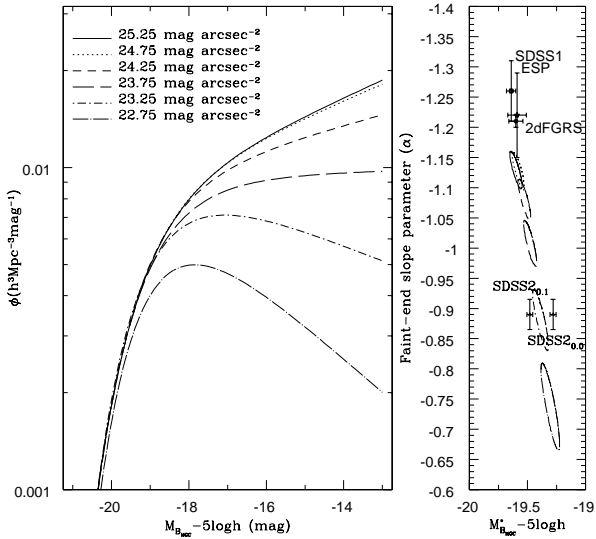} 
\caption{The left panel shows the result of a re-analysis of the MGC
data with various low surface brightness selection limits imposed (as
indicated). This figure demonstrates the extreme sensitivity of the
recovered Schechter function parameters to the spectroscopic surface
brightness completeness limit. The right panel shows $1$-$\sigma$
error ellipses for the Schechter function fits along with the data
points from the right panel of Fig.\ \ref{lfcomp}.}
\label{lfcomp2}
\end{figure}

\subsubsection{Exploring the discrepancy with the SDSS LFs} 
\label{sdssdiff}
The most noticeable outliers in Fig.~\ref{lfcomp} are the two SDSS
results, SDSS1 \citep{blanton01} and SDSS2 \citep{blanton03}. The
discrepancy in the normalisation of SDSS1 has been comprehensively
discussed by both \cite{norberg02} and \cite{blanton03} and ascribed
to the lack of an evolutionary model in the SDSS1 analysis, coupled
with their method of normalisation. An over-density in the SDSS
commissioning data region also contributes. The SDSS1 normalisation
has been revised by \cite{mgc1}: $\phi^*_{\mbox{\tiny \sc
SDSS1}}(\mbox{NEW}) = 0.76 \, \phi^*_{\mbox{\tiny \sc
SDSS1}}(\mbox{OLD}$). Once corrected, the SDSS1 LF agrees well with
the MGC (2$\sigma$-discrepancy in $\alpha$), the 2dFGRS and the ESO
Slice Project (ESP) estimates.

The more recent SDSS2 result has a significantly fainter
$M^*$\footnote{Note that the conversion from $^{0.1}\!g$ to $\bmgc$ is
given by \cite{blanton03} as $\bmgc= \, ^{0.1}\!g-0.09$.} and a much
flatter faint-end slope than any of the other surveys.
\cite{blanton03} argued that the difference in the $r$-band LFs of
SDSS1 and SDSS2 was simply due to the inclusion of luminosity
evolution in the latter and that the luminosity densities measured by
the 2dFGRS and SDSS2 agreed to within the errors once the differing
amounts of evolution had been taken into account. However, in the
$g$-band the discrepancy in $\alpha$ between SDSS1 and SDSS2 is more
than twice that in the $r$-band. Furthermore, even though the
luminosity evolution derived in the SDSS2 fitting process is very
strong ($Q = 2.04$~mag per unit redshift, where $M^*(z) = M^*(z=0) - Q
\, z$) it cannot explain away the difference between the MGC and SDSS2
values. Our evolutionary correction given in Section~\ref{evolution}
corresponds to $Q \approx 0.74$. If we replace this with the SDSS2
evolution, $E(z) = -2.04 \, z$, and re-derive our collapsed MGC LF we
find $M_B^* - 5 \log h = -19.36$~mag and $\alpha =-1.08$ (or $M_B^* -
5 \log h = -19.36$~mag and $\alpha =-1.06$ if we restrict the LF
fitting to the equivalent SDSS2 limit of $M_B - 5 \log h < -16$
mag). Hence some tension remains between the SDSS2 and MGC LFs
irrespective of evolution. The same argument holds when comparing
SDSS2 to the 2dFGRS since the latter adopted an evolution equivalent
to $Q\approx 1.15$.

To test whether the surface brightness selection limits might be
responsible, Fig.\ \ref{lfcomp2} shows a re-analysis of the MGC data,
following the same procedures as before, but now imposing
progressively shallower surface brightness limits (as indicated). The
left panel again shows the LFs and the right panel the $1$-$\sigma$
error ellipses of the Schechter function fits. It is quite apparent
that shallower surveys progressively miss a greater fraction of the
dwarf population(s) -- and ultimately giants -- recovering both a
fainter $M^*$ and a more positive $\alpha$. In fact, Fig.\
\ref{lfcomp2} based on MGC data mirrors the expectation from the
simulations of \cite{cross02} remarkably well (see their Fig.~5).  

On Fig.\ \ref{lfcomp2} the LF with an imposed selection limit of
$23.25$\mpas\ appears to match the SDSS2 LF closely and initially
suggests that surface brightness selection might explain the SDSS2
result. (Note that SDSS1 implemented a BBD-style analysis). However
Fig.\ \ref{completeness} (left panels) do not show a significant
difference in the low surface brightness spectroscopic completeness
between the 2dFGRS and SDSS surveys suggesting limits closer to
$24.25$\mpas. We note that the SDSS2 surface brightness $50$ per cent
completeness limit (including consideration of deblending and flux
loss issues, etc.) is quoted as $\mu^e_r[{\rm Circular}] \approx
23.4$\mpas\ (\citealp{blanton04}, their fig.\ 3), where $[{\rm
Circular}]$ indicates that the SDSS surface brightness measures are
derived from `circular'\footnote{The SDSS2 surface brightness measure
is derived from the combination of SDSS Petrosian magnitudes and S{\'
e}rsic half-light radii based on circular aperture photometry. At this
point it is not clear why the SDSS team have adopted circular
apertures and how one should interpret circular half-light radii for
moderate and highly inclined systems, particularly given the
additional dependencies this will introduce as a function of both
seeing and distance. It is also not clear why \citet{blanton03} prefer
S{\' e}rsic fitted half-light radii rather than their more reliable
(\citealp{blanton04b}) empirically measured Petrosian radii (which
agree closely with MGC circularised half-light radii).} half-light
radii. Conversely the MGC (and most contemporary datasets) use major
axis half-light radii. These are derived by growing the optimal
ellipse until it encloses $50$ per cent of the flux. The use of a
circular aperture will always result in a lower half-light radius than
an elliptical aperture (except for purely face-on systems), and hence
in a higher effective surface brightness. The two effective surface
brightnesses are related by:
\begin{equation}
\label{sbcorrection}
\mu^e[{\rm Major Axis}] = \mu^e[{\rm Circular}]-2.5\log
\left(\frac{b}{a} \right),
\end{equation}
where $(b/a)$ is the axis ratio. 
For a random distribution of infinitely thin disks the expected median
axis ratio $\langle b/a \rangle$ is $0.5$. In reality the galaxy
distribution is more diverse consisting of spheroids, bulges and thick
disk components/systems. Circular half-light radii measurements are
also somewhat resolution dependent, introducing both seeing and
redshift dependencies into the conversion. Empirically we find for our
MGC data: $\langle b/a \rangle_{\rm MGC}=0.71$, implying a correction
of $0.37$\mpas. Although we cannot directly compare our MGC elliptical
half-light radii to the circular S{\' e}rsic half-light radii used by
SDSS2, we can compare them to the standard SDSS-DR1 circular Petrosian
half-light radii. We find $\langle r_{50,{\rm Petrosian}} / r_{\rm
MGC} \rangle = 0.84$, independently confirming the correction of
$0.37$\mpas. Hence the quoted SDSS2 $50$ per cent completeness limit
of $\mu^e_r[{\rm Circular}]=23.4$\mpas\ equates to $\mu^e_r[{\rm Major
Axis}]=23.77$\mpas.


The colour correction is more complex as it depends strongly on galaxy
type and luminosity. The low luminosity population is the bluest
($\langle \bmgc-r \rangle \approx 0.5$~mag) and hence surface
brightness incompleteness will preferentially affect the faint end of
the LF. Evidence for such a colour completeness bias can be seen in
the observed SDSS incompleteness within the MGC region. This is shown
as a function of $(g-r)$ or $(u-r)$ versus $\mu^e_{\rm app}$ in the
middle row (centre and right panels) of Fig.~\ref{completeness}. As
the SDSS spectroscopic survey is $r$-band selected it should not be
surprising that the $\bmgc$ completeness is a more complex surface in
magnitude, surface brightness and colour. In the worst case scenario
we find a 50 per cent completeness limit for the bluest systems of
$\mu^e_{\bmgc} = 24.27$\mpas\ which is consistent with
Fig.~\ref{completeness} (middle left), but more importantly
significantly fainter than the surface brightness limit implied by
Fig.~\ref{lfcomp2}. That a colour selection is obvious in both the
$(g-r)$ and $(u-r)$ SDSS completeness maps of Fig.~\ref{completeness}
argues that the colour selection bias is not a feature of the $u$ band
data but more a general trend across all filters.

Adopting an appropriate $\mu^e_{\bmgc}$ selection criterion is clearly
non-trivial, however a simple way forward is to construct a LF from
the available SDSS2 redshifts within the MGC region (see
Table. 1). This will have the SDSS selection function inherently
built in. We adopt $m_{\mbox{\tiny \sc faint}} = 18.5$~mag,
$\mu^e_{\mbox{\tiny \sc lim}} = 24.27$\mpas, the SDSS2 $g$-band
evolutionary correction and apply a conventional monovariate step-wise
maximum likelihood \citep{eep}. We now recover $M^*_B=(-19.26 \pm
0.07)$~mag and $\alpha=-0.83 \pm 0.06$ (or $M^*_B=(-19.26 \pm 0.07)$~mag
and $\alpha=-0.82 \pm 0.06$ for $M_B < -16$~mag) and consistent at the
$\sim 1$-$\sigma$ level with SDSS2. We therefore conclude
that the discrepancy between the MGC and SDSS2 LFs is complex, but
predominantly due to the colour bias within the SDSS
(cf.\ Fig.~\ref{completeness} middle right). This highlights the
additional complexity in measuring LFs for filters other than the
spectroscopic selection filter.

\subsubsection{Lower limits on the very faint-end}
\label{llfe}
\begin{figure}
\centering\includegraphics[width=\columnwidth]{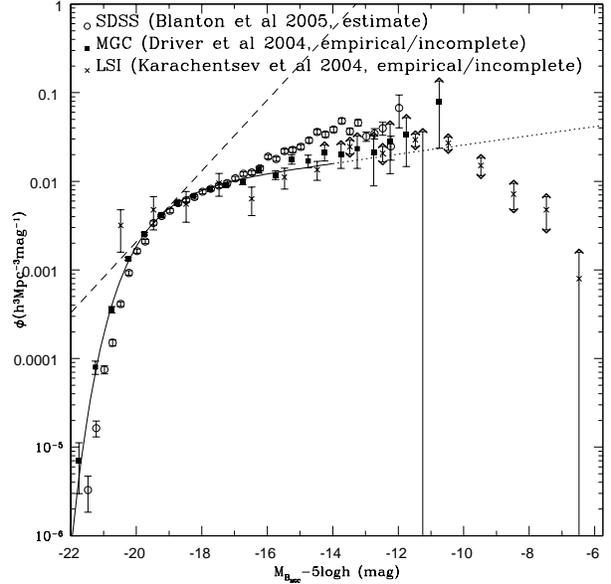}
\caption{Tentative constraints on the very faint end of the galaxy
luminosity function from three independent studies. Solid squares:
available MGC data using the BBD-style analysis but with all limits
removed. These data are incomplete at the faint end with the loss of
both high and low surface brightness systems and therefore represent
lower limits only. Open circles: a recent `speculative' SDSS estimate
\citep{blanton04} based on an extrapolation of the surface brightness
distribution at bright magnitudes. Crosses: Local Sphere of Influence
($D<5$~Mpc) data taken from the catalogue of \citet{karach04}. The
Schechter function from Fig.\ \ref{lfcomp} is shown as the solid line
and its extrapolation as the dotted line. The dashed line shows the
contour of equal contribution to the overall luminosity density.}
\label{guesstimate}
\end{figure}

Finally we note the recent preprint by \citet{blanton04} in which an
attempt is made to sample the very low luminosity regime from their
New York University Value Added Catalogue \citep{blanton04b}. This
incorporates a strategy for correcting the surface brightness
incompleteness within the SDSS2 dataset resulting in a significantly
steeper faint-end slope ($\alpha = -1.28$ to $-1.52$ in $r$). The
premise is that the surface brightness distribution of the dwarf
population can be extrapolated from that of the giant population. This
results in correction factors for the space density of dwarf systems
of up to $3.5$ (see \citealp{blanton04}, their fig.~6). In a similar
spirit of producing a `best guess' LF at low luminosities Fig.\
\ref{guesstimate} shows the MGC LF with the selection boundaries and
lower redshift bounds removed. This undoubtedly represents an
underestimate of the true space density at low luminosities
($M_B -5 \log h > -14$~mag), but in so far as a comparison can
be made it is consistent with the SDSS result \citep{blanton04}, in
the sense that the SDSS data lie above the MGC lower limits.  Our LF
is also consistent with a compilation of $231$ galaxies within the
local sphere ($D<5$~Mpc) taken from \citet{karach04}. Apart from
indicating the amount of work that is yet to be done, Fig.\
\ref{guesstimate} does suggest that the low redshift global luminosity
function must steepen beyond $\alpha=-1.13$ at some intermediate
magnitude ($M_B - 5\log h > -16$~mag). However it would seem
unlikely that it could steepen enough for the dwarf systems to make a
significant contribution to the total local luminosity density (as
indicated by the dashed line).

\subsection{The surface brightness distribution}
\label{sbd}

\begin{figure*}
\centering\includegraphics[width=\textwidth]{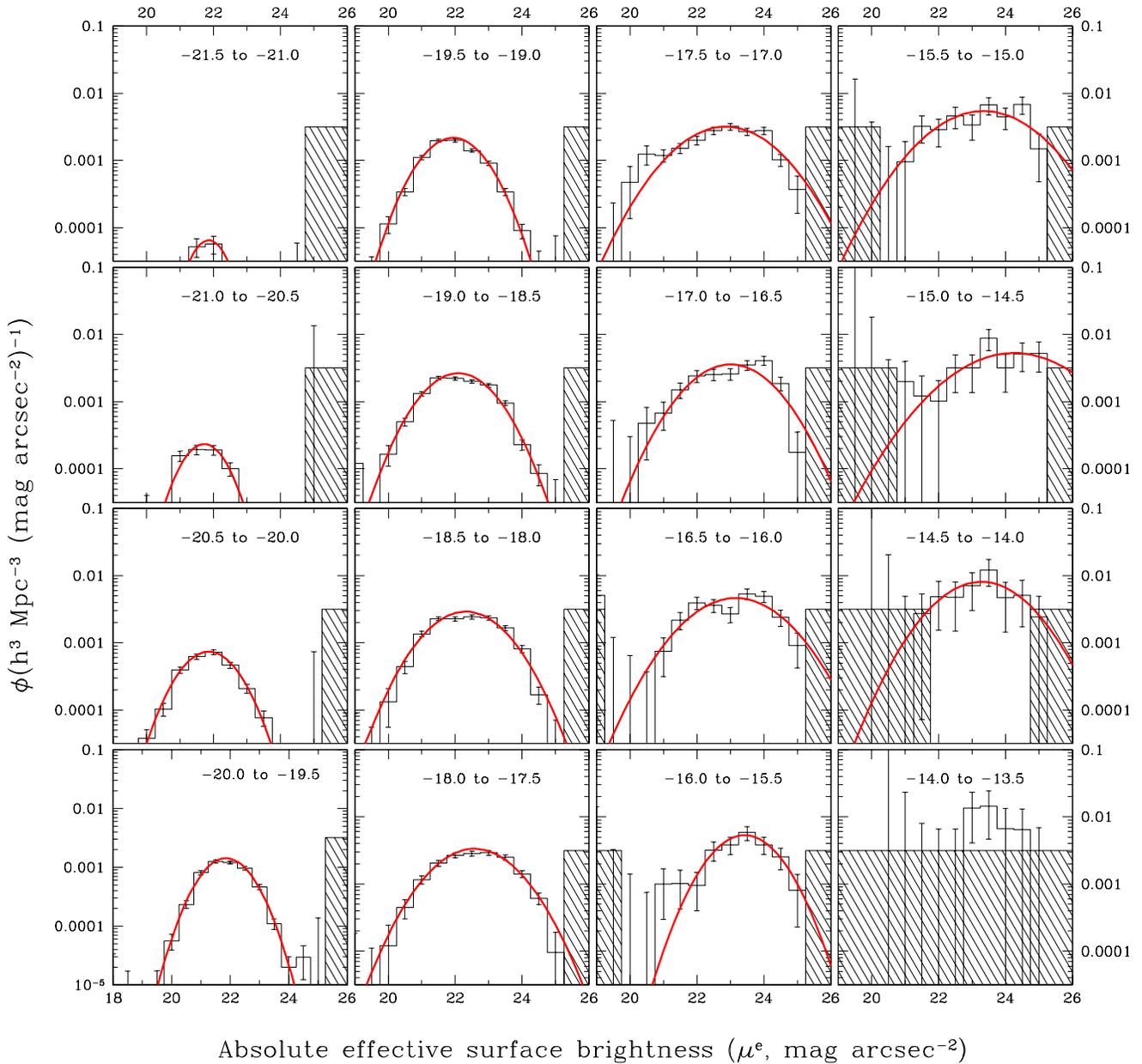} 
\caption{Slices of the BBD along the surface brightness axis in
intervals of constant luminosity as indicated (histogram with
errorbars). Note that the data within shaded regions are not used in
the determination of the Gaussian fits (shown as solid lines).}
\label{sbcomp}
\end{figure*}

As we found earlier the Cho\l oniewski function adopted by previous
studies (\citealp{dejong}; \citealp{cross02}) provides a poor fit to
the joint luminosity-surface brightness distribution. This is mainly
due to the apparent broadening of the surface brightness distribution
towards fainter luminosities. Fig.~\ref{sbcomp} highlights this by
showing the surface brightness distribution at progressively fainter
absolute magnitude intervals along with the selection limits taken
from Fig.~\ref{bbdcont}. We attempt to fit each of these distributions
(Fig.~\ref{sbcomp}, solid lines) with a Gaussian function defined
as:
\begin{equation}
\phi(\mu^e)=\frac{\phi^*}{\sqrt{2\pi} \sigma_{\mu^e}}
\exp\left[-\frac{1}{2}\left(\frac{\mu^{e*}-\mu^e}{\sigma_{\mu^e}}
\right)^2\right],
\end{equation}
where, $\mu^{e*}$ is the characteristic (peak) surface brightness and
$\sigma_{\mu^e}$ is the dispersion (analogous to those values in the
Cho\l oniewski function). The errors reflect $\sqrt{n}$-statistics
from the original observed number of galaxies contributing to each
histogram element. For elements with zero values (i.e., no galaxies
detected) we adopt the error appropriate for a detection of one galaxy
so that the significance of these `null' detections can be used to
help constrain the fits. The number of degrees of freedom (listed in
the final column of Table~\ref{gfdata}) is the number of data values
plus any zero values either side of the distribution within the
selection boundaries minus the number of fitting parameters
(three). Fig.~\ref{sbgauss}(a) shows the $\mu^{e*}$ (dark grey line)
and 1-$\sigma_{\mu^e}$ values (shaded region) as a function of
absolute magnitude. These results are also tabulated in
Table~\ref{gfdata}. Fig.~\ref{sbgauss}(a) shows a trend of an
initially invariant luminosity-surface brightness relation until $M_B
- 5 \log h \approx -19$~mag, and then a steady decline to the limits
of detection at $M_B - 5 \log h \approx -14$~mag. The dispersion of
the surface brightness distributions grows steadily broader as the
luminosity decreases. The changes in both the luminosity-surface
brightness relation and its dispersion explain the poor fit of the
Cho\l oniewski function which cannot accommodate such features.

At this stage it is also worth noting that the Gaussian fits return
poor $\chi^2_{\mbox{\tiny min}}$ values at intermediate luminosities
with even a hint of bimodality (at $M_B - 5 \log h \approx
-16$~mag). This may imply that the low luminosity distribution
consists of two overlapping populations: rotationally supported disks
(lower surface brightnesses) and pressure supported spheroids (higher
surface brightnesses). If borne out by more detailed structural
studies then one might expect the rotationally supported disk systems
to exhibit both a steeper $M$-$\mu$ relation and a narrower surface
brightness distribution than the overall population.

\subsubsection{Comparison to established structural trends}
Three luminosity-surface brightness trends ($L$-$\Sigma$) are well
known (see Table~\ref{lfdata}): the Kormendy relation for spheroids
(\citealp{kormendy}); Freeman's Law for disks \citep{ken}; and the
Virgo/Local Group dwarf relation (\citealp{impey88}; \citealp{mateo}).
More recently \citet{dejong} identified a relation for late-type disks
similar to that seen for dwarf systems in Virgo and the Local Group.
The general trends for these established relations are overlain on
Fig.~\ref{sbgauss}(a). One can speculate that the combination of these
trends might well lead to the overall shape of the global
$L$-$\Sigma$-distribution shown on Fig.~\ref{sbgauss}(a), as the
spheroid-to-disk ratio typically declines with decreasing luminosity.
Hence the global trend initially follows a Kormendy relation,
switching to a more constant Freeman Law and finally into a declining
dwarf relation. In comparison to the \citet{dejong} sample for
late-type disks we find a shallower luminosity-surface brightness
relation and a broader surface brightness distribution. This may have
two reasons: firstly, the dJL sample is $I$-band selected and our use
of a global filter conversion may introduce this error; secondly, the
dJL sample is far more local and both inclination and
dust-corrected. An obvious extension to the MGC analysis would be to
also implement these corrections.

Obviously, detailed structural analysis is required before a
meaningful comparison to these ad hoc local studies can be
made. However, we can highlight two key questions: (i) Are there two
distinct disk populations, i.e.\ a giant disk population adhering to
Freeman's Law and a dwarf-disk population which does not, or just a
single dwarf-disk relation? (ii) Is the dwarf surface brightness
distribution bimodal reflecting the two dominant classes of dwarf
galaxies (dwarf ellipticals and dwarf irregulars)?

\subsubsection{Comparison to previous global $L$-$\Sigma$ studies}
There are three significant studies of the global $L$-$\Sigma$
distribution (see Table~\ref{lfdata}): a 2dFGRS study by
\citet{cross01}, an SDSS study by \citet{shen03} and a local
galaxy SDSS study by \citet{blanton04}.

%
%
%

\begin{figure*}
\centering\includegraphics[width=\textwidth]{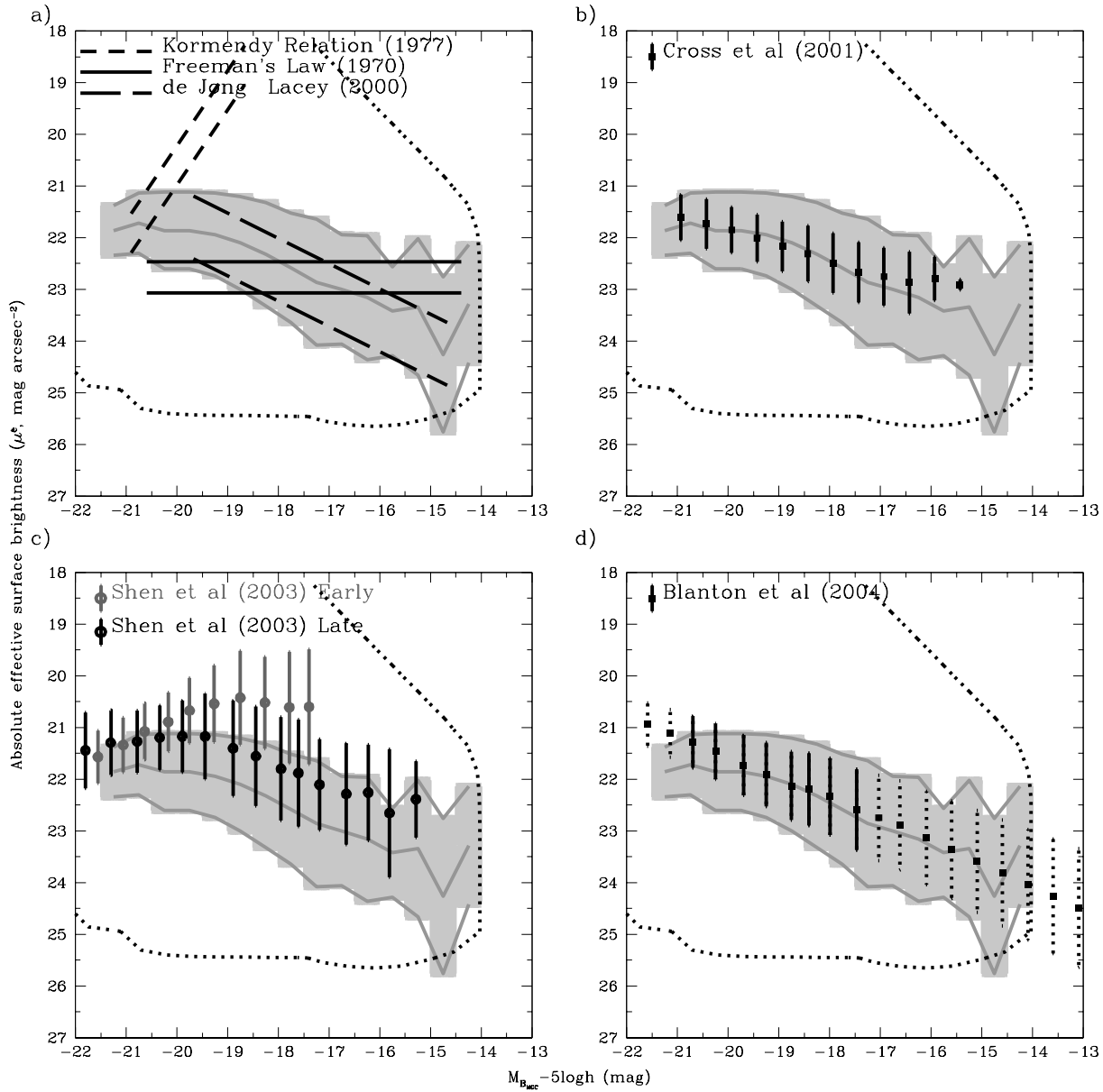}
\caption{Four panels showing the 1-$\sigma_{\mu^e}$ ranges (grey
shading) and the $\mu^{e*}$ values (dark grey line) from the Gaussian
fits of Fig.~\ref{sbcomp}. On all panels the MGC selection boundary is
shown by the outer dotted line and represents a volume limit of
1800~$h^{-3}$Mpc$^3$, (see Fig.\ \ref{bbdcont}). (a) Overlain are the
$1$-$\sigma$ ranges for the canonical spheroid (Kormendy relation; short
dashed line), disk (Freeman's Law; solid horizontal line) and
late-type (\citealp{dejong}; long dashed line) results. (b) A
comparison to the $b_J$-band 2dFGRS results of \citet{cross01},
data points and $1$-$\sigma$ ranges. (c) A comparison to the $r$-band
SDSS results of \citet{shen03}. Details of the colour correction
are given in the text. Shen et al.\ divide their sample by
concentration index into `early' ($c > 2.86$, black symbols) and
`late' types ($c < 2.86$, grey symbols). At most magnitudes the
late-types dominate the space density and so the most meaningful
comparison, band-switch not withstanding, is between the MGC data
points and the SDSS late types. (d) A comparison between the global
SDSS2 distribution \citep{blanton04} transformed from $r$ to $\bmgc$
as for (c). The dashed symbols indicate the region of extrapolation.}
\label{sbgauss}
\end{figure*}

In comparison to \citet{cross01} we find that the MGC data follow a
similar $L$-$\Sigma$ relation but show significantly broader surface
brightness distributions. Fig.~\ref{sbgauss}(b) shows the comparison
where we have taken the data from Table C2 of \citet{cross01} and
fitted Gaussian distributions in a similar manner as for the MGC
data. The obvious discrepancy is in the $\sigma_{\mu^e}$ values and
this reflects a number of significant improvements. Firstly, the
deeper MGC isophote coupled with the improved redshift completeness of
MGCz (see Fig.~\ref{completeness}) enables the inclusion of lower
surface brightness systems. As these preferentially occur at lower
luminosity the effect acts to broaden the surface brightness
distribution at fainter luminosities. Secondly, the 2dFGRS sample was
not seeing-corrected and hence the sizes of the high surface
brightness galaxies were overestimated. Thirdly, the Cross et al.\
analysis derived the surface brightness measurements under the
assumption that all galaxies were perfect exponential disks. These
factors conspire to push the 2dFGRS BBD into a narrower
distribution. A fourth factor, hard to qualitatively assess, is the
major revisions to the 2dFGRS photometry which have taken place since
\citet{cross01} (see \citealp{mgc3} for further details on the 2dFGRS
recalibration history). 
 
In Fig.~\ref{sbgauss}(c) we overlay the recent SDSS results by
\citet{shen03}. The Shen et al.\ study is $r_{0.1}$-band selected with
an imposed surface brightness cut at $\mu^e_{r}[{\rm
Circular}]=23$\mpas. Shen et al.\ divide their sample into a number of
sub-groupings and we elect to show the data from their Fig.~13 which
separates galaxies according to concentration index, $c$ (see
\citealp{nakamura} for details). This divides the population into
galaxies earlier or later then S0/a (with an $83$ per cent success
rate). To convert the Shen et al.\ $r$-band results to $\bmgc$ we
derive the early and late-type $\langle \bmgc-r \rangle$ colour in
each of the $r$-band absolute magnitude bins specified by Shen et al.
To do this we assume our eyeball E/S0 type corresponds to `early' and
our eyeball Sabcd/Irr to `late' (a detailed account of the eyeball
morphologies and morphological luminosity functions is presented in
Driver et al., in preparation). Hence each data point is adjusted
independently. In addition we must also correct the SDSS effective
surface brightness measurements from circular to major axis
measurements as discussed in Section \ref{sdssdiff}, i.e.\ $\mu^e[{\rm
Major Axis}] = \mu^e[{\rm Circular}]+0.37$. Fig.~\ref{sbgauss}(c)
shows the comparison between the Shen et al. data and the MGC and we
see a uniform offset of $\delta \mu^e \approx 0.4$ \mpas. Despite
extensive experimentation with various data subsets and selection
boundaries we have been unable to reproduce this result from the MGC
data. 

Fig.~\ref{sbgauss}(d) shows a comparison between the recent very low
redshift SDSS $r$ results \citep{blanton04} and the MGC. We implement
the same colour and surface brightness corrections as for the Shen et
al.\ data. The ridge line of the very local SDSS data shows good
agreement with the MGC data, albeit with a narrower distribution
(i.e., lower $\sigma$ values). However this appears to be a further
manifestation of the SDSS use of circular apertures. For example, if
we derive the surface brightness distribution around the $M^*$-point
for the MGC data using circular and major axis derived surface
brightness measures we find a peak offset of $0.42 \pm 0.03$ (slightly
larger than our estimated median offset of $0.37$) with $\sigma_{\rm
Major Axis} = 0.74 \pm 0.02$ and $\sigma_{\rm Circular} = 0.60 \pm
0.02$. This latter value is now consistent with the recent SDSS
estimate (\citealp{blanton04}). Note that the dashed lines on
Fig.~\ref{sbgauss} indicated the SDSS extrapolated data used to
correct the faint-end of the luminosity function. To the MGC limit
these extrapolated distributions appear to match our data well
providing some vindication for the extrapolation process (see
\citealp{blanton04}, Section \ref{llfe} and Fig.~\ref{guesstimate}).

The above two comparison between the MGC and the \cite{shen03} and
\cite{blanton04} studies must imply an inconsistency between the two
SDSS results. This is perhaps more obvious if one bypasses the MGC
data and directly compare Fig.~14 of Shen et al with Fig.~5 of Blanton
et al. The peaks of the late-type/low sersic index SB distributions
(both measured in $r$) are clearly distinct. Given the consistency
with the 2dFGRS and the traditional structural studies discussed
earlier we conclude that the problem most likely lies with the Shen et
al. analysis. If we disregard the Shen et al study, we can conclude
that the ridge line of the effective surface brightness distribution
is well constrained and consistent between the MGC, 2dFGRS and the
most recent SDSS results. However the MGC distribution is
significantly broader than either the 2dFGRS or SDSS results which is
due to improvements in the analysis (in comparison to the 2dFGRS) and
the use of elliptical as opposed to circular aperture photometry (in
comparison to the SDSS).

\subsection{Constraining the evolution of $M^*$ galaxies}
\label{evolmstar}
\begin{figure}
\centering\includegraphics[width=\columnwidth]{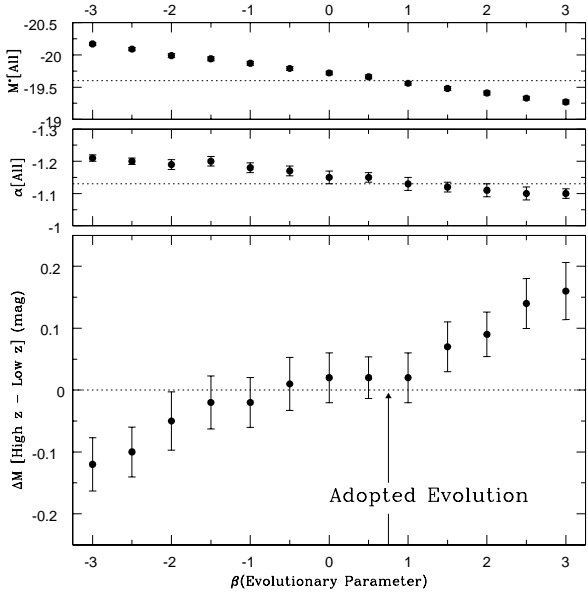}
\caption{Main panel: constraining the evolutionary parameter $\beta$
from the variation of $M^*$ between low and high redshift
samples. Upper panels: the $M^*$ and $\alpha$ values for the entire
dataset for different beta values.}
\label{beta}
\end{figure}

In Section \ref{evolution} we described our adopted evolutionary
parameter, $\beta$, where the luminosity evolution follows the form:
$E(z)=2.5\log[(1+z)^{-\beta}]$. In Section \ref{sdssdiff} we saw that
this evolution differs from that adopted by the 2dFGRS and SDSS2 and
discussed the possible impact. Fig.~\ref{beta} (upper panels)
highlights this dependency by showing the derived values for $M^*$ and
$\alpha$ for various values of $\beta$. Taking this one step further
we can attempt to constrain $\beta$ directly by adopting the
additional constraint that the recovered $M^*$ parameter should be
invariant to redshift. We divide our sample into two redshift
intervals: $0.013 < z < 0.1$ and $0.1 < z < 0.18$ (with median
redshifts of $0.08$ and $0.13$ respectively) and redetermine the
collapsed LF distributions using our BBD analysis for regular
intervals of $\beta$. We hold $\alpha$ fixed to the value derived from
the full redshift range for each $\beta$ (i.e., the value indicated in
the middle panel of Fig.\ \ref{beta}) and solve for $M^*$ and
$\phi^*$. It is necessary to hold $\alpha$ fixed because the high-$z$
sample has too few low-luminosity galaxies to adequately constrain it.
Fig.\ \ref{beta} (main panel) shows the resulting $\Delta M^*$ versus
$\beta$ and we find the weak constraint: $-2.0 < \beta < 1.25$. While
this is not particularly stringent, it is consistent with our adopted
value of $\beta = 0.75$. This constraint on beta is marginally
inconsistent ($2 \sigma$) with the evolution found by
\citet{blanton03} for the $g$-band ($\beta \approx 2$), but consistent
with the evolution adopted by the 2dFGRS ($\beta \approx 1$). Perhaps
more importantly this range of $\beta$ implies a potential systematic
uncertainty on the quoted $M^*$ ($\alpha$) values of $\pm 0.25$ ($\pm
0.04$). As noted at the start of Section \ref{cosmoke}, it is also
equally plausible that this error does not actually reflect evolution
per se but could instead be interpreted as a redshift dependent
photometric error or similar.

\section{Conclusions}
\label{conclusions}
In this paper we have attempted to recover the space density and
surface brightness distributions of galaxies, paying careful attention
to surface brightness selection biases. We have used the Millennium
Galaxy Catalogue which, although smaller in size, has significantly
higher resolution, greater depth and higher redshift completeness than
either the 2dFGRS or SDSS datasets, hence probing to fainter intrinsic
luminosities. In comparison to most earlier studies this work includes
the following enhancements.

~

\noindent
1. Imaging data which probes $\sim 1$--$2$\mpas\ deeper
   than the 2dFGRS and SDSS surveys.

\noindent
2. Exceptionally high spectroscopic completeness ($96$ per cent).

\noindent
3. Individual $K$-corrections derived from $uBgriz$ fits to spectral
   templates.

\noindent
4. Seeing-corrected half-light radius measurements without assumption
   of profile shape.

\noindent
5. Modeling of the MGC surface brightness detection and reliability
   limits which includes the effects of the analysis software.

\noindent
6. A weighting system which accounts for apparent magnitude and
   apparent surface brightness dependent redshift incompleteness.

\noindent
7. A joint luminosity-surface brightness step-wise maximum likelihood
   method which incorporates the selection boundaries.

\subsection{The luminosity function}
\label{lfresult}
Having incorporated these additional improvements to the classical
measurement of galaxy luminosity functions, we generally find close
agreement with previous results. In particular, our collapsed
luminosity distribution is marginally higher in normalisation and
marginally flatter than the 2dFGRS \citep{norberg02}, ESP \citep{esp}
and SDSS1 (\citealp{blanton01}; after renormalisation by
\citealp{mgc1}) results. It is significantly brighter and steeper than
the more recent SDSS2 luminosity function \citep{blanton03}. We infer
that this discrepancy arises because of a colour-selection bias in the
SDSS. The MGC survey extends to lower luminosity than these previous
surveys, providing more leverage and hence more reliability at the
faint end. Our final Schechter function parameters are: $\phi^* =
(0.0177 \pm 0.0015) \, h^3$~Mpc$^{-3}$, $M_B^* - 5 \log h = (-19.60
\pm 0.04)$~mag and $\alpha =-1.13 \pm 0.02$ with a $\chi^2/\nu$ of
$19.6/15$ indicating a respectable fit. Overall we find that surface
brightness selection effects do not play a significant role at bright
luminosities. However, the combination of the luminosity-surface
brightness relation and the broadening of the surface brightness
distribution is likely to lead to the loss of both low surface
brightness and compact dwarf systems in contemporary surveys -- hence
the variation in $\alpha$ (cf.\ Fig.~\ref{lfcomp} \&~\ref{lfcomp2} and
discussion in Section \ref{spacedensity}). At the very faint end we
provide lower limits which suggest that the luminosity function must
turn up at some point faintwards of $M_{\bmgc} - 5 \log h \approx
-16$~mag.

\subsection{The low redshift $b_J$ luminosity density}
The final MGC Schechter function parameters, which include
compensation for the effects of selection bias, imply a luminosity
density of $j_{b_J}=(1.99 \pm 0.17) \times 10^8 \, h
\,L_{\odot}$~Mpc$^{-3}$ (adopting $M_{\odot} = +5.3$ mag and
$\bmgc=b_{J}-0.131$, cf.\ \citealp{mgc1}). This is fully consistent
with the MGC revised values for the 2dFGRS, ESP and SDSS1 surveys (see
Table 3 of \citealp{mgc1}). It is also worth noting that this value is
also consistent with the original 2dFGRS (\citealp{norberg02}) and ESP
(\citealp{esp}) values. The more recent SDSS2 result
(\citealp{blanton03}) finds $j_{b_j}^{z=0}=(1.54 \pm 0.10) \times
10^8 \, h \, L_{\odot}$~Mpc$^{-3}$ and $j_{b_j}^{z=0.1}=(1.90 \pm
0.10) \times 10^8 \, h \, L_{\odot}$~Mpc$^{-3}$. Hence while the
$z=0.1$ value is consistent the $z=0$ value is not ($3 \sigma$).
The SDSS $z=0$ value is derived by adopting the evolutionary
parameters derived by \cite{blanton03}. If we use our evolution
instead ($\beta=0.75$, see Section~\ref{evolution}) the SDSS
$z=0$ luminosity density becomes $j_{b_j}^{z=0}=(1.72 \pm 0.10)
\times 10^8 \, h \, L_{\odot}$~Mpc$^{-3}$ which although closer is
also inconsistent with our result. From Section~\ref{sdssdiff} we
concluded that a colour bias may have resulted in an underestimate of
the SDSS $g$-band LF and hence to an underestimate of the SDSS
luminosity densities.

~

\noindent Based on the consistency between the MGC, 2dFGRS and ESP
results we can also find that while the more sophisticated BBD
analysis does recover some additional flux from the low surface
brightness population, it does not constitute a significant change
(consistent with the findings of \citealp{driver99}).

\subsection{The joint luminosity-surface brightness relation}
Our recovered bivariate brightness distribution exhibits a clear
`ridge' of data well separated from the selection boundaries down to
$M_B - 5 \log h \approx -16$~mag at which point both very compact and
highly extended dwarf systems will be missed. This indicates that the
recovered Schechter function is generally robust to surface brightness
selection effects for luminous galaxies but that the dwarf population(s)
remain elusive. We attempted to fit a Cho\l oniewski function
\citep{chol}, essentially a Schechter function with a Gaussian
distribution in surface brightness, to the overall distribution. It
introduces three additional parameters: a characteristic surface
brightness; a Gaussian dispersion; and a luminosity surface brightness
relation. We find that this function fails to provide a satisfactory
fit due to the broadening of the surface brightness distribution with
decreasing luminosity and a distinct change in slope of the
luminosity-surface brightness relation at $M_B - 5 \log h \approx
-19$~mag.

\subsection{The surface brightness distribution(s)}
We find that the surface brightness distribution of galaxies exhibits
interesting behaviour, with a well bounded Gaussian-like distribution
at $L^*$, qualitatively consistent with the combination of the
Kormendy relation for spheroids and a $2\times$ broader Freeman's Law
for bright disks. At $L^*$ we find that the surface brightness
distribution is well described by a Gaussian with parameters: $\phi^*
= (3.5 \pm 0.1) \times 10^{-2} h^3$~Mpc$^{-3}$, $\mu^{e*} = (21.90 \pm
0.01)$\mpas, $\sigma_{\mu^e} = (0.77 \pm 0.02)$\mpas\ with a
$\chi^2/\nu$ of $9.8/8$. At fainter luminosities the normalisation
increases, the characteristic surface brightness becomes fainter and
the dispersion broadens until the distribution appears almost flat (or
possibly bimodal). The Schechter function fit to the Gaussian
normalisations (column 4 of Table 3) provides a basic completeness
correction and yields consistent Schechter function parameters to
those quoted in Section \ref{lfresult}, i.e.\ $\phi^* = (0.0184 \pm
0.0008) \, h^3$~Mpc$^{-3}$, $M_B^* - 5 \log h = (-19.55 \pm 0.08)$~mag
and $\alpha =-1.11 \pm 0.02$ with a $\chi^2/\nu$ of $26.0/12$.

In comparison to the 2dFGRS study of \citet{cross01} we find a
consistent luminosity-surface brightness relation but with a
significantly broader dispersion in surface brightness. This is
predominantly due to the implementation of seeing corrected size,
direct (i.e., profile-independent) half-light radius measurements and
improved completeness for extended systems. A comparison to the SDSS
results of \cite{shen03} and \cite{blanton04} are complex because of
the SDSS standard of circular aperture photometry. However we find
that our results can be reconciled with those of \cite{blanton04} but
not with those of \cite{shen03}. We also note that the SDSS use of
circular apertures also leads to significantly narrower surface
brightness distributions.

\subsection{Confrontation with theory}
Various studies (\citealp{fall}; \citealp{dss97}; \citealp{mo};
\citealp{dejong}; \citealp{bell03}) relate the Gaussian dispersion of
the surface brightness distribution, or equivalently of the
$R_e$\footnote{$R_e$ is the half-light radius measured in kpc. It is
related to surface brightness by $\mu^e = M + 2.5 \log (2\pi R_e^2) +
36.57$ and hence $\sigma_{\mu^e} = 2.17 \sigma_{\ln R_e}$.}
distribution, to the dimensionless spin parameter of the dark matter
halo, $\lambda = J|E|^{\frac{1}{2}}G^{-1}M^{-\frac{5}{2}}_{\mbox{\tiny
halo}}$ \citep{peebles}. The basic concept is that the baryons in the
disk reflect the angular momentum state (rotational or dispersive) of
the dark matter halo. While the luminosity relates to the mass, the
disk scale-length relates to the specific angular momentum via the
spin parameter. Numerical simulations generally predict values for
$\sigma_{\ln R_e} \approx 0.56 \pm 0.04$ (\citealp{bullock}; see also
\citealp{cole}). This is most consistent with our observations at
fainter luminosities. However, at brighter luminosities we find a
significantly narrower dispersion $\sigma_{\ln R_e} = 0.36 \pm
0.01$. This may imply that brighter, more mature galaxies have evolved
via mechanisms currently beyond the standard numerical simulations
(i.e., the baryons have decoupled from the dark matter halo and
evolved away from the halo properties). Alternatively it may argue for
significant non-directional baryon infall onto the halo, muddying any
coupling between the primordial baryons and the halo
\citep{katz}. Further possibilities might also include the presence of
bulges in high luminosity systems whose properties depend more
critically on a central supermassive black hole \citep{silk}.

One intriguing result of the simulations is the lack of variation of
the surface brightness or size distribution with mass
(\cite{vitvitska}. However, the more detailed studies which actually
follow individual halos {\em do} imply a greater variation in
$\lambda$ for low-mass systems. For example, \cite{vitvitska} trace
the history of three halos through the merging process. They clearly
show that $\lambda$ initially varies significantly as the halos grow
but that it becomes more resilient to mergers once the halos reach
sufficient mass. Ultimately all three halos converge towards a common
value at high mass. This seems plausible as one would expect smaller
systems to be more affected by mergers than more massive systems.

\subsection{Summary}
The surface brightness distribution of galaxies has been hotly debated
for over thirty years since the advent of Freeman's Law and the
conjecture that it is wholly a selection bias \citep{disney}.  While
in many minds the debate is resolved through deep HI studies (e.g.\
\citealp{deephi}) and Ly-$\alpha$ absorption line studies
(\citealp{mo}; \citealp{lya}), others remain loyal to Disney's
conjecture (e.g.\ O'Neil, Andreon \& Culliandre 2003) and still argue
for a flat surface brightness distribution at all luminosities
(\citealp{spray}; \citealp{dalcanton}; \citealp{dss97}) implying a
significant amount of hidden mass. At least in O'Neil et al.'s case
this stance derives from neglecting the clear luminosity-surface
brightness relation seen in their data. The global surface brightness
distribution may indeed be flat but the low surface brightness systems
are predominantly dwarfs, providing little enhancement to the global
luminosity, baryon or mass densities. We have attempted to address this
debate through careful consideration of {\em most} selection
issues. We probe to depths where the distribution is clearly bounded
and where we robustly demonstrate we could reliably recover giant low
surface brightness galaxies were they to exist in significant numbers.
They do not.

While we address the selection bias in the redshift survey as a
function of apparent magnitude and apparent surface brightness other
potential selection effects remain unexplored. For example, the
redshift completeness may well be a function of spectral type, in
which case the assumption that all galaxies of similar apparent
magnitude and surface brightness are drawn from the same redshift
distribution will not be correct. In particular, inert low surface
brightness galaxies which may have insufficient features to readily
yield a redshift and may be missed entirely. We also do not at this
stage decompose galaxies into bulge and disk components. This is
clearly required by both the observations (which show two, or possibly
three, distinct luminosity-surface brightness relations for spheroid
and disk components) as well as by theory and simulations which argue
for distinct formation mechanisms for these components (merger and
accretion respectively; e.g.\ Pierani, Mohayaee \& de Freitas Pacheco
2004). We hence intend to extend our analysis in three ways: $100$ per
cent spectroscopic completeness; bulge-disk decompositions; and
extension to longer wavelengths and in particular the near-IR which
promises smoother, dust-free profiles.

\section*{ACKNOWLEDGMENTS}
We thank Shiyin Shen, Eric Bell and Michael Blanton for useful
discussions and exchange of data, Steve Phillipps, Alister Graham and
Stefan Andreon for comments on the early drafts and the referee for a
number of detailed scientific and stylistic improvements.

The Millennium Galaxy Catalogue consists of imaging data from the
Isaac Newton Telescope and spectroscopic data from the Anglo
Australian Telescope, the ANU 2.3m, the ESO New Technology Telescope,
the Telescopio Nazionale Galileo, and the Gemini Telescope. The survey
has been supported through grants from the Particle Physics and
Astronomy Research Council (UK) and the Australian Research Council
(AUS). The data and data products are publicly available from
http://www.eso.org/$\sim$jliske/mgc/ or on request from J. Liske or
S.P. Driver.

This research has made use of the NASA/IPAC Extragalactic Database
(NED) which is operated by the Jet Propulsion Laboratory, California
Institute of Technology, under contract with the National Aeronautics
and Space Administration.

This research made use of mock catalogues constructed by the Durham
astrophysics group and the VIRGO consortium.

Funding for the Sloan Digital Sky Survey (SDSS) has been provided by
the Alfred P.~Sloan Foundation, the Participating Institutions, the
National Aeronautics and Space Administration, the National Science
Foundation, the U.S.\ Department of Energy, the Japanese
Monbukagakusho, and the Max Planck Society. The SDSS Web site is
http://www.sdss.org/. The SDSS is managed by the Astrophysical
Research Consortium (ARC) for the Participating Institutions. The
Participating Institutions are The University of Chicago, Fermilab,
the Institute for Advanced Study, the Japan Participation Group, The
Johns Hopkins University, Los Alamos National Laboratory, the
Max-Planck-Institute for Astronomy (MPIA), the Max-Planck-Institute
for Astrophysics (MPA), New Mexico State University, University of
Pittsburgh, Princeton University, the United States Naval Observatory,
and the University of Washington.

\onecolumn

\pagebreak

\section*{}

\begin{table*}
\begin{minipage}{9.5cm}
\caption{The breakdown of the survey contributions to the final MGC
redshift sample.}
\begin{tabular}{lrrl} 
\hline
Survey & No.\ of galaxies & No.\ of best & Reference\\ 
& with $Q_z \ge 3$ & redshifts \\
\hline
MGCz & $5054$ & $4944$ & This work\\
SDSS-DR1 & $2909$ & $1528$ & \cite{abazajian03} \\
2dFGRS & $4129$ & $3152$ & \cite{2dfgrs}\\
2QZ & $19$ & $13$ & \cite{2qz}\\
PF & $37$ & $28$ & \cite{francis04}\\
LSBG & $11$ & $2$ & \cite{impey96}\\
NED & $1200$ & $29$ & NED\\
\hline
\end{tabular}
\label{alldata}
\end{minipage}
\end{table*}

\begin{table*}
\begin{minipage}{\textwidth}
\caption{Cho\l oniewski (upper) and Schechter (lower) function
parameters for the MGC galaxy distribution and comparable parameters
from other surveys corrected to the $\bmgc$ bandpass. MGC BBD refers
to the empirical MGC BBD integrated over surface brightness.  Values
are for $\Omega_M = 0.3, \Omega_{\Lambda}=0.7$ and
$H_0=100$~km~s$^{-1}$.}
\begin{tabular}{llccccccl} \hline
Survey & Filter & $\phi^*$ & $M^*_{\bmgc}$ & $\alpha$ &
$\mu^{e*}$ & $\beta$ & $\sigma_{\mu^e}$ & Ref \\ & & $10^{-2} \,
h^3$~Mpc$^{-3}$ & mag & & \mpas & & \mpas & \\ \hline MGC (All types)
& $B_{\mbox{\tiny \sc MGC}}$ & $2.13 \pm 0.32$ & $-19.37 \pm 0.15$ &
$-0.99 \pm 0.14$ & $22.0 \pm 0.13$ & $0.33 \pm 0.15$ & $0.86 \pm 0.09$
& 1\\ 
Spheriods & $B$ & --- & $-19.70^a$ & --- & $20.03$ & $-1.57$ & $0.38$ & 2 \\
Disks & $B$     & --- & $-19.70^a$ & --- & $22.77$ & $0.0$ & $0.3$ & 3\\
HDF (All types) & $B_{AB}+0.13$ & --- & $-19.70^a$ & --- & $20.9 \pm
0.8$ & $0.67 \pm 0.09$ & $~1$ & 4\\ E/S0s (Total) & $R+1.4$ & $0.010 \pm
0.002$ & $-20.7 \pm 0.2$ & $1.06 \pm 0.26$ & $20.8 \pm 0.4$ & $0.92
\pm 0.09$ & $-0.62 \pm 0.08$ & 5 \\ E/S0s (Bulges) & $R+1.4$ & $0.008
\pm 0.003$ & $-20.1 \pm 0.3$ & $1.30 \pm 0.50$ & $20.32 \pm 0.35$ &
$0.84 \pm 0.09$ & $-0.55 \pm 0.10$ & 5 \\ Sd/Irrs(Total) & $I_C+1.8$ &
$0.332 \pm 0.07$ & $-19.7 \pm 0.1$ & $-0.93 \pm 0.10$ & $21.8 \pm 0.2$
& $0.51 \pm0.04$ & $0.61 \pm 0.04$ & 6 \\ Sd/Irrs(Disk) & $I_C+1.8$ &
$0.332 \pm 0.07$ & $-19.9 \pm 0.1$ & $-0.90 \pm 0.10$ & $21.7 \pm 0.2$
& $0.43 \pm0.05$ & $0.78 \pm 0.07$ & 6 \\ 2dFGRS & $b_{\mbox{\tiny
2dF}}+0.06$ & $2.06 \pm 0.09$ & $-19.67 \pm 0.02$ & $-1.05 \pm 0.04$ &
$21.90 \pm 0.02$ & $0.28 \pm 0.02$ & $0.57 \pm 0.01$ & 7 \\
SDSS(Early) & $r_{0.1}+1.3$ & --- & $-19.70^a$ & --- & $19.4$ & $1.2$ &
$0.6-1.2$ & 8 \\ SDSS(Late) & $r_{0.1}+0.9$ & --- & $-19.70^a$ & --- &
$21.35$ & $0.54$ & $0.6-1.0$ & 8 \\ \hline MGC BBD & $B_{\mbox{\tiny
\sc MGC}}$ & $1.77 \pm 0.15$ & $-19.60 \pm 0.04$ & $-1.13\pm 0.02$ &
--- & --- & --- & 1 \\ ESP & $b_{\mbox{\tiny \sc J}} +0.13$ & $1.65
\pm 0.3$ & $-19.59 \pm 0.08$ & $-1.22 \pm 0.07$ & --- & --- & --- & 9
\\ SDSS1 & $g+0.40$ & $2.06 \pm 0.23$ & $-19.64 \pm 0.04$ & $-1.26 \pm
0.05$ & --- & --- & --- & 10 \\ 2dFGRS & $b_{\mbox{\tiny 2dF}}+0.06$ &
$1.61 \pm 0.06$ & $-19.60\pm0.008$ & $-1.21 \pm 0.01$ & --- & --- &
--- & 11\\ SDSS2$_{0.0}$ & $^{0.1}g-0.09$ & $2.12 \pm 0.08$ & $-19.28
\pm 0.02$ & $-0.89 \pm 0.03$ & --- & --- & --- & 12 \\ SDSS2$_{0.1}$ &
$^{0.1}g-0.09$ & $2.18 \pm 0.08$ & $-19.48 \pm 0.02$ & $-0.89 \pm
0.03$ & --- & --- & --- & 12 \\ \hline
\end{tabular}
\label{lfdata}
$^a$Adopted value.\\
References: (1) This work. (2) \cite{kormendy}. (3) \cite{ken}. (4) \cite{driver99}. (5)
\cite{dejong04}. (6) \cite{dejong}. (7) \cite{cross01}. (8) \cite{shen03}. (9)
\cite{esp}. (10) \cite{blanton01}. (11) \cite{norberg02}. 
(12) \cite{blanton03} with and without evolution corrections.
\end{minipage}
\end{table*}

\begin{table*}
\begin{minipage}{10.5cm}
\caption{Gaussian fits to the surface brightness distribution in
various intervals of absolute magnitude. Surface brightness is defined
as the mean surface brightness within the half-light semi-major axis
(i.e., using elliptical apertures).}
\begin{tabular}{ccccc} \hline
$M_{\bmgc} - 5 \log h$ & $\mu^{e*}$ & $\sigma_{\mu^e}$
& $\phi^*$ & $\chi^2/\nu$ \\
mag & \mpas & \mpas & $10^{-3} h^3$~Mpc$^{-3}$ & \\ \hline
$-21.25$&$21.86\pm0.10$&$0.48\pm0.12$&$0.08\pm0.02$&$4.0/5$\\
$-20.75$&$21.72\pm0.04$&$0.58\pm0.03$&$0.34\pm0.03$&$14.0/10$\\
$-20.25$&$21.86\pm0.02$&$0.74\pm0.01$&$1.37\pm0.06$&$5.67/10$\\
$-19.75$&$21.86\pm0.01$&$0.74\pm0.01$&$2.64\pm0.09$&$10.5/9$\\
$-19.25$&$21.94\pm0.01$&$0.80\pm0.01$&$4.33\pm0.10$&$16.4/10$\\
$-18.75$&$22.10\pm0.02$&$0.90\pm0.01$&$5.96\pm0.13$&$30.5/11$\\
$-18.25$&$22.32\pm0.03$&$1.00\pm0.01$&$7.29\pm0.22$&$27.6/9$\\
$-17.75$&$22.58\pm0.03$&$1.06\pm0.02$&$8.86\pm0.27$&$ 7.61/9$\\
$-17.25$&$22.86\pm0.07$&$1.22\pm0.07$&$9.69\pm0.65$&$16.6/9$\\
$-16.75$&$23.00\pm0.09$&$1.06\pm0.07$&$9.48\pm0.68$&$22.7/8$\\
$-16.25$&$23.16\pm0.12$&$1.20\pm0.10$&$13.9\pm0.9$&$16.0/7$\\
$-15.75$&$23.42\pm0.11$&$0.86\pm0.12$&$11.5\pm2.1$&$4.12/7$\\
$-15.25$&$23.34\pm0.24$&$1.32\pm0.20$&$17.9\pm3.0$&$6.85/7$\\
$-14.75$&$24.26\pm0.43$&$1.50\pm0.16$&$19.8\pm3.7$&$4.31/6$\\
$-14.25$&$23.28\pm0.64$&$1.14\pm0.40$&$23.0\pm7.9$&$1.38/3$\\ \hline
\end{tabular}
\label{gfdata}
\end{minipage}
\end{table*}

\pagebreak

\appendix

\section{Simulations}
\label{appsim}
Galaxy sampling, detection and analysis selection effects are
extremely complex (see e.g.\ Disney, Davies \& Phillipps 1989) and
depend on: galaxy profile; inclination; opacity; detector
characteristics; survey depth and resolution; and the detection and
analysis process, in particular background smoothing and removal.  In
this appendix we intend to crudely establish the empirical selection
limits of the MGC through simulations. The equations defining the
redshift dependent selection limits are described in
Section~\ref{sellim} (see also \citealp{driver99}) and require a
maximum and minimum size limit and a limiting surface brightness.
In addition, we need a method to correct the observed half-light radii
of compact galaxies for the effects of seeing.

\begin{figure*}
\centering\includegraphics[width=\textwidth]{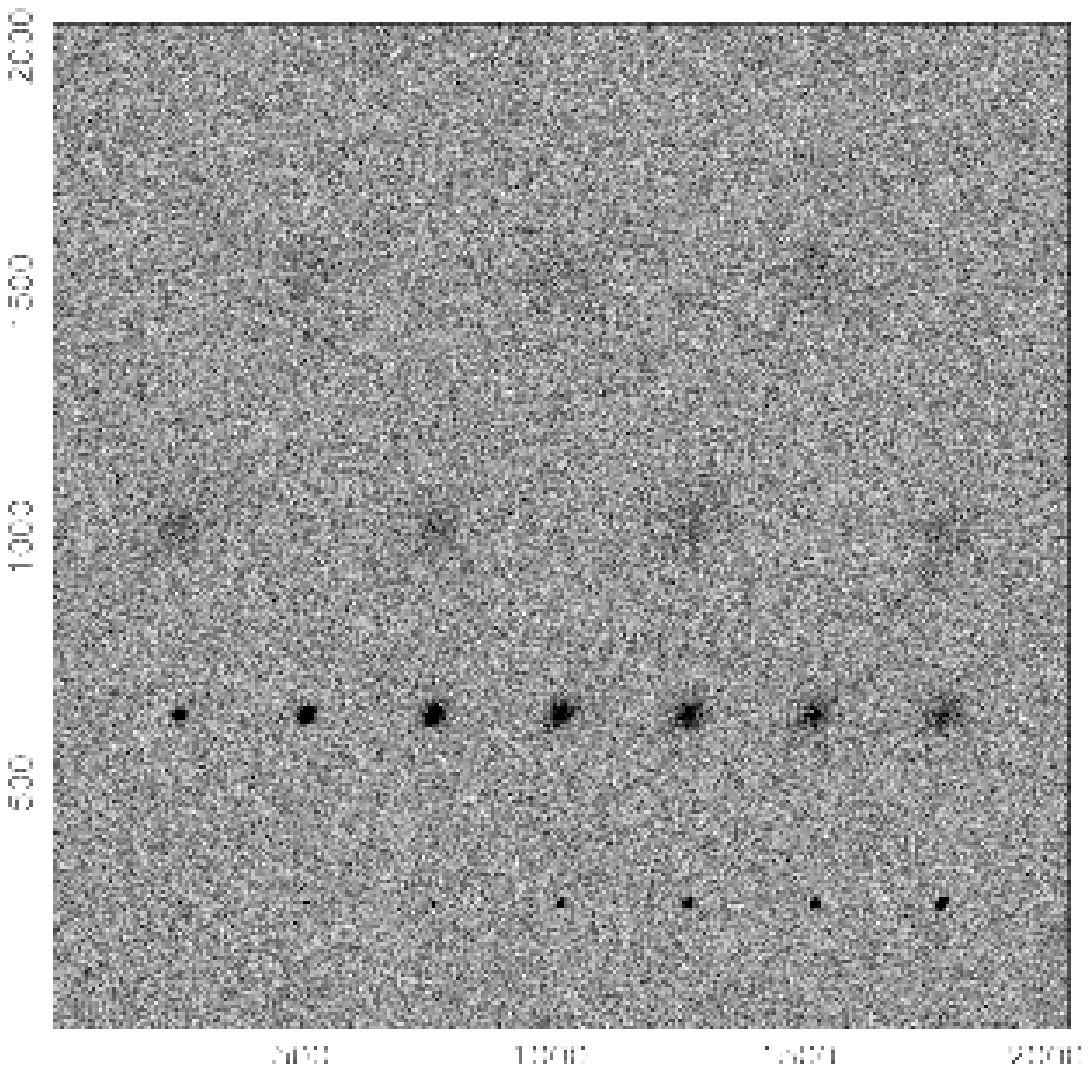}
\caption{A  simulated  image  showing $45^\circ$-inclined  exponential
profile  galaxies all  with  magnitude $18$~mag,  but with  half-light
radii  varying  from  $0.1$  to $32$~arcsec  (from  left-to-right  and
bottom-to-top $0.1, 0.2, 0.4, 0.8, 1.0,  1.2, 1.5, 2, 4, 6, 8, 10, 12,
15, 20, 22, 24,  26, 28, 30, 32$~arcsec). X and Y  units are in pixels
($\equiv 0.333$~arcsec).  
Figure degraded see http://www.eso.org/$\sim$jliske/mgc/ for full pdf
copy of this paper.}
\label{simulations}
\end{figure*}

To establish the selection limits and the correction procedure we use
{\sc iraf}'s {\sc artdata} package to simulate $45^\circ$-inclined
optically thin exponential disks (this is consistent with our
definition of surface brightness, see Section \ref{mgcsb}) on a CCD
frame with identical pixel size, sky noise and seeing as the MGC
imaging data (i.e., $0.333$~arcsec/pixel, $\mu_{\mbox{\tiny \sc
lim}}=26$\mpas\ and $\Gamma = 1.27$~arcsec). We produce simulation
plates at $0.5$~mag intervals from $m_{\mbox{\tiny in}} = 16$ to
$20$~mag, each containing galaxies with half-light radii varying from
$r_{\mbox{\tiny in}} = 0.1$ to $32$~arcsec (see
Fig.~\ref{simulations}). Compact galaxies were also simulated with de
Vaucouleurs profiles but the results were the same as those for
exponential disks.

The MGC catalogue was derived using the SExtractor analysis software
\citep{sex}. The full details of this process are described in
\cite{mgc1}. We use SExtractor in an identical manner (e.g., sky
background mesh $256\times256$ pixels with $7\times7$ filtering within
each mesh) on the simulated data and also measure half-light radii in
the same way as for the real data.

Tables~A1-A3 contain the results of our simulations as a function of
$m_{\mbox{\tiny in}}$ and $r_{\mbox{\tiny in}}$ and these are also
presented in Fig.\ \ref{selection}.

\subsection{Seeing correction}
\begin{figure*}
\centering\includegraphics[width=\textwidth]{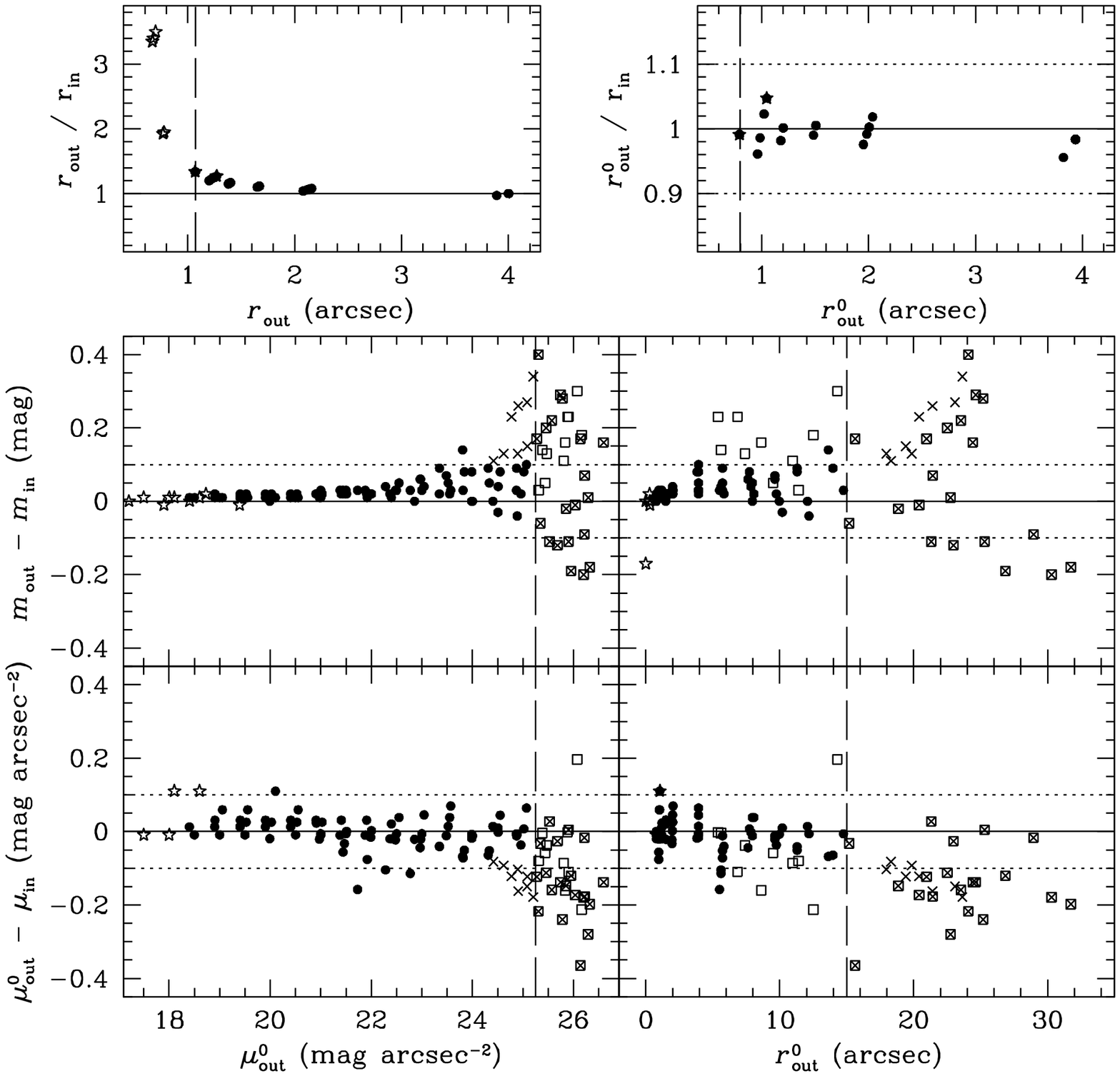}
\caption{A graphical representation of Tables A1, A2 and A3. The top
panels show the ratio of the output and input half-light radii both
with (right) and without (left) seeing correction. These panels
demonstrate the effectiveness of the seeing correction. The dashed
lines show the minimum size limit $r_{\mbox{\tiny \sc min}} = 0.6 \,
\Gamma + 0.31$~arcsec. The middle panels show the difference between
the output and input magnitudes as a function of the seeing-corrected
output half-light radius ($r^0_{\mbox{\tiny out}}$, right) and as a
function of the seeing-corrected output effective surface brightness
($\mu^0_{\mbox{\tiny out}}$, left). Finally, the bottom panels show
difference between the seeing-corrected output and input surface
brightnesses as a function of the same two parameters. The dashed
lines in the left panels show the low surface brightness limit
$\mu^0_{\mbox{\tiny \sc lim}} = 25.25$\mpas, while in the right panels
they mark the maximum size limit $r^0_{\mbox{\tiny \sc max}} =
15$~arcsec.  In all panels we show objects identified as stars by
SExtractor as open star symbols, objects beyond $r^0_{\mbox{\tiny \sc
max}}$ as crosses and objects beyond $\mu^0_{\mbox{\tiny \sc lim}}$ as
open squares. Solid symbols represent galaxies within the selection
limits and for these we recover magnitudes and surface brightnesses
that lie consistently within $\pm 0.1$~mag and $\pm 0.1$\mpas\ of
their input values.}

\label{selection}
\end{figure*}

From the top left panel of Fig.~\ref{selection} we can see the impact
of the seeing, which was set to the median value of our survey. To
crudely correct for this effect we make the resonable assumption that
the observed half-light radius, $r_e$, is given by the quadratic sum
of the true half-light radius, $r^0_e$, and some fraction, $\eta$, of
the seeing FWHM, $\Gamma$:
\begin{equation}
\label{hlrcorr}
r_e^2 = (r^0_e)^2 + \eta \, \Gamma^2.
\end{equation}
From the real MGC data we find that the median $r_e$ of stars is well
described by $0.6 \, \Gamma$ (for a perfectly Gaussian PSF this would
be $0.5 \, \Gamma$) and hence one would expect $\eta = 0.36$. However,
from the simulations we find $\eta = 0.32$ and that
equation~(\ref{hlrcorr}) recovers the true half-light radii to an
accuracy of better than $10$ per cent for objects not classified as
stars (see top right panel of Fig.~\ref{selection}).

\subsection{Minimum size limit}
The most compact simulated galaxies are routinely classified as stars
by the SExtractor star-galaxy separation and these objects are marked
with an asterisk in Table~A1. SExtractor's ability to distinguish a
galaxy of given $r_e$ from stars cannot primarily depend on the
absolute value of $r_e$. It must rather depend on the galaxy's
distance, $\Delta$, from the stellar locus in $r_e$-space, where the
scale of that distance is set by the width of the stellar $r_e$
distribution. From the MGC data we find the width of that distribution
to be independent of seeing. Hence the minimum $\Delta$ at which a
galaxy can still be reliably distinguished from stars cannot depend on
seeing either. Recalling that the position of the stellar locus is
given by $0.6 \, \Gamma$ we thus find:
\begin{equation}
\label{rmin}
r_{\mbox{\tiny \sc min}} = 0.6 \, \Gamma + \Delta_{\mbox{\tiny \sc min}}
\end{equation}
To determine $\Delta_{\mbox{\tiny \sc min}}$ we note from Table~A1
that in the simulations the confusion between stars and galaxies sets
in at $r_{\mbox{\tiny \sc min}}(\Gamma = 1.27) = 1.07$~arcsec,
implying $\Delta_{\mbox{\tiny \sc min}} = 0.31$~arcsec. The same value
is found by fitting the actually observed minimum $r_e$ of galaxies in
each MGC field as a function of $\Gamma$.

Finally, combining equations (\ref{hlrcorr}) and (\ref{rmin}) we find the
minimum seeing-corrected half-light radius:
\begin{equation}
r^0_{\mbox{\tiny \sc min}} = \sqrt{0.04 \, \Gamma^2 + 0.37 \, \Gamma +
  0.1}
\end{equation}

\subsection{Maximum size and low surface brightness limits}
Here we define these limits (on the seeing-corrected quantities) not as
absolute detection limits but rather as the limits to which we can
expect `reliable' photometry. We arbitrarily define `reliable' to mean
magnitude and surface brightness estimates accurate to $0.1$~mag and
$0.1$\mpas\ respectively, thereby limiting the potential systematics
due to isophotal errors and background subtraction to $0.1$~mag and
$0.1$\mpas. As most of our analysis uses $0.5$~mag and
$0.5$\mpas-sized bins this ensures minimal impact of systematics on
our final BBD.

To determine $r^0_{\mbox{\tiny \sc max}}$ and $\mu^0_{\mbox{\tiny \sc
lim}}$ we proceed in the following way: using Tables A2 and A3 we find
for a given $m_{\mbox{\tiny in}}$ the smallest $r_{\mbox{\tiny in}}$
for which either the recovered magnitude, $m_{\mbox{\tiny out}}$, or
the recovered (seeing-corrected) surface brightness
$\mu^0_{\mbox{\tiny out}}$ differs by at least $0.1$~mag or
$0.1$\mpas\ from the respective input value. Given the resulting pair
of ($m_{\mbox{\tiny in}}$, $r_{\mbox{\tiny in}}$) values we again use
Tables A2 and A3 to find the corresponding ($m_{\mbox{\tiny out}}$,
$\mu^0_{\mbox{\tiny out}}$) pair and plot this on Fig.~\ref{maxhlr}
(lower square points with errorbars). Finally, we find those values of
$r^0_e$ and surface brightness which best fit these points. We find
\begin{equation}
r^0_{\mbox{\tiny \sc max}}= 15 \mbox{~arcsec}
\end{equation}
and
\begin{equation}
\mu^0_{\mbox{\tiny \sc lim}}= 25.25 \mbox{\mpas}.
\end{equation}

Finally it is worth noting that some of the extreme low surface
brightness galaxies were detected by SExtractor but they were broken
up into numerous small objects. Rerunning SExtractor over the entire
MGC with the deblender switched off may enable us to eventually probe
fainter than the limits determined here, however a detailed study of
the MGC using a cross-correlation method \citep{roberts} failed to
find any additional galaxies.

\begin{table*}
\begin{minipage}{11.5cm}
\caption{The recovered half-light radii ($r_{\mbox{\tiny out}}$, uncorrected
for seeing) from our simulations. Objects marked with a $^*$ were
classified as stars by SExtractor.}
\begin{tabular}{c|ccccccccc}
\hline
$r_{\mbox{\tiny in}}$ & \multicolumn{9}{c}{Simulated magnitude
  ($m_{\mbox{\tiny in}}$)} \\ 
arcsec & 16.0 & 16.5 & 17.0 & 17.5 & 18.0 & 18.5 & 19.0 & 19.5 & 20.0 \\ \hline
0.1& 0.63*& 0.63*& 0.63*& 0.64*& 0.64*& 0.64*&0.67*&0.66*&0.68*\\	    
0.2& 0.67*& 0.67*& 0.67*& 0.67*& 0.67*& 0.67*&0.67*&0.68*&0.70*\\	    
0.4& 0.77*& 0.77*& 0.77*& 0.78*& 0.77*& 0.78*&0.78*&0.77*&0.78*\\	    
0.8& 1.07*& 1.07*& 1.07 & 1.07 & 1.07 & 1.07 &1.07 &1.07 &1.07 \\	    
1.0& 1.27*& 1.27*& 1.25 & 1.25 & 1.27 & 1.25 &1.22 &1.20 &1.20 \\	    
1.2& 1.40 & 1.40 & 1.40 & 1.40 & 1.40 & 1.40 &1.38 &1.38 &1.38 \\	    
1.5& 1.67 & 1.67 & 1.67 & 1.67 & 1.67 & 1.67 &1.67 &1.67 &1.65 \\	    
  2& 2.13 & 2.13 & 2.13 & 2.13 & 2.08 & 2.11 &2.08 &2.13 &2.16 \\	    
  4& 4.00 & 4.00 & 4.00 & 4.00 & 4.00 & 4.00 &3.89 &4.00 &4.00 \\	    
  6& 5.55 & 5.66 & 5.66 & 5.88 & 5.77 & 5.77 &5.77 &5.66 &5.44 \\	    
  8& 7.99 & 7.66 & 8.10 & 7.99 & 7.88 & 7.77 &7.44 &6.88 \\	    
 10& 9.66 & 9.66 & 9.99 &10.21 & 9.77 & 9.54 &8.66 \\	    
 12&11.32 &11.32 &12.10 &12.21 &11.43 &10.99 \\	    
 15&13.65 &13.99 &14.77 &15.21 &14.32 &12.54 \\	    
 20&18.32 &17.98 &21.32 &18.87 &15.65 \\		    
 22&19.87 &19.42 &22.98 &20.42 \\		    
 24&20.42 &20.98 &25.31 &21.43 \\			    
 26&21.42 &22.53 &26.86 &22.76 \\			    
 28&23.09 &23.53 &28.97 &24.42 \\			    
 30&23.64 &24.64 &30.30 \\			    
 32&24.09 &25.20 &31.75 \\
\hline                                
\end{tabular}
\end{minipage}
\end{table*}

\begin{table*}
\begin{minipage}{11.5cm}
\caption{The recovered magnitudes ($m_{\mbox{\tiny out}}$) from our
simulations.}
\begin{tabular}{c|ccccccccc}
\hline
$r_{\mbox{\tiny in}}$ & \multicolumn{9}{c}{Simulated magnitude
  ($m_{\mbox{\tiny in}}$)} \\
arcsec & 16.0 & 16.5 & 17.0 & 17.5 & 18.0 & 18.5 & 19.0 & 19.5 & 20.0 \\ \hline
0.1&16.00&16.50&17.00&17.50&18.00&18.50&19.00&19.50&20.00\\
0.2&16.00&16.50&17.00&17.50&18.00&18.33&19.00&19.50&20.00\\
0.4&16.00&16.50&17.00&17.50&18.00&18.49&19.00&19.52&19.99\\
0.8&16.01&16.51&17.01&17.51&18.01&18.50&19.01&19.51&20.02\\
1.0&16.01&16.51&17.01&17.51&18.01&18.51&19.01&19.53&20.01\\
1.2&16.01&16.51&17.01&17.51&18.01&18.52&19.03&19.53&20.02\\
1.5&16.02&16.52&17.02&17.52&18.02&18.52&19.02&19.51&20.00\\
2  &16.02&16.52&17.02&17.52&18.02&18.52&19.03&19.54&20.03\\
4  &16.03&16.53&17.02&17.53&18.03&18.55&19.08&19.58&20.10\\
6  &16.03&16.54&17.03&17.52&18.03&18.55&19.09&19.64&20.23\\
8  &16.05&16.56&17.02&17.50&18.04&18.58&19.13&19.73\\
10 &16.06&16.57&17.00&17.47&18.02&18.55&19.16\\
12 &16.09&16.58&17.00&17.46&18.03&18.61\\
15 &16.14&16.59&17.03&17.44&18.30&18.68\\
20 &16.11&16.63&16.89&17.48&18.17\\
22 &16.13&16.65&16.88&17.49&12.13\\
24 &16.23&16.67&16.89&17.57\\
26 &16.26&16.70&16.81&17.51\\
28 &16.27&16.72&16.91&17.66\\
30 &16.34&16.79&16.80\\
32 &16.40&16.78&16.82\\
\hline   
\end{tabular}
\end{minipage}
\end{table*}

\begin{table*}
\caption{The input ($\mu_{\mbox{\tiny in}}$, left) and
  seeing-corrected output ($\mu^0_{\mbox{\tiny out}}$, right)
  effective surface brightnesses from our simulations. Objects
  classified as stars are indicated by ***** and objects not detected
  are indicated by ---.}
\begin{tabular}{c|ccccccccc}
\hline
$r_{\mbox{\tiny in}}$ & \multicolumn{9}{c}{Simulated magnitude
  ($m_{\mbox{\tiny in}}$)} \\
arcsec & 16.0    & 16.5     & 17.0     & 17.5     & 18.0     & 18.5     & 19.0     & 19.5     & 20.0 \\ \hline
0.1 &13.00 *****& 13.50 *****& 14.00 *****& 14.50 *****& 15.00 *****& 15.50 *****& 16.00 *****& 16.50 *****& 17.00 *****\\	     
0.2 &14.50 *****& 15.00 *****& 15.50 *****& 16.00 *****& 16.50 *****& 17.00 *****& 17.50 *****& 18.00 *****& 18.50 *****\\	     
0.4 &16.01 *****& 16.51 *****& 17.01 *****& 17.51 *****& 18.01 *****& 18.51 *****& 19.01 *****& 19.51 *****& 20.01 *****\\	     
0.8 &17.51 *****& 18.01 *****& 18.51 18.50& 19.01 19.00& 19.51 19.50& 20.01 19.99& 20.51 20.50& 21.01 21.00& 21.51 21.51\\	     
1.0 &18.00 *****& 18.50 *****& 19.00 19.06& 19.50 19.56& 20.00 20.11& 20.50 20.56& 21.00 20.98& 21.50 21.44& 22.00 21.92\\	     
1.2 &18.39 18.40& 18.89 18.90& 19.39 19.40& 19.89 19.90& 20.39 20.40& 20.89 20.91& 21.39 21.38& 21.89 21.88& 22.39 22.37\\	     
1.5 &18.88 18.91& 19.38 19.41& 19.88 19.91& 20.38 20.41& 20.88 20.91& 21.38 21.41& 21.88 21.91& 22.38 22.40& 22.88 22.85\\	     
2   &19.50 19.53& 20.00 20.03& 20.50 20.53& 21.00 21.03& 21.50 21.47& 22.00 22.00& 22.50 22.48& 23.00 23.05& 23.50 23.57\\	     
4   &21.01 21.00& 21.51 21.50& 22.01 21.99& 22.51 22.50& 23.01 23.00& 23.51 23.52& 24.01 23.99& 24.51 24.55& 25.01 25.07\\	     
6   &21.89 21.73& 22.39 22.28& 22.89 22.77& 23.39 23.35& 23.89 23.81& 24.39 24.33& 24.89 24.87& 25.39 25.38& 25.89 25.88\\	     
8   &22.51 22.55& 23.01 22.97& 23.51 23.55& 24.01 24.00& 24.51 24.51& 25.01 25.02& 25.51 25.47& 26.01 25.90& 26.51 -----\\	     
10  &23.00 22.97& 23.50 23.48& 24.00 23.99& 24.50 24.51& 25.00 24.96& 25.50 25.44& 26.00 25.84& 26.50 -----& 27.00 -----\\	     
12  &23.39 23.35& 23.89 23.84& 24.39 24.41& 24.89 24.89& 25.39 25.31& 25.89 25.81& 26.39 -----& 26.89 -----& 27.39 -----\\	     
15  &23.88 23.81& 24.38 24.31& 24.88 24.87& 25.38 25.34& 25.88 26.07& 26.38 26.16& 26.88 -----& 27.38 -----& 27.88 -----\\	     
20  &24.50 24.42& 25.00 24.90& 25.50 25.53& 26.00 25.85& 26.50 26.14& 27.00 -----& 27.50 -----& 28.00 -----& 28.50 -----\\	     
22  &24.71 24.61& 25.21 25.09& 25.71 25.68& 26.21 26.03& 26.71 -----& 27.21 -----& 27.71 -----& 28.21 -----& 28.71 -----\\	     
24  &24.90 24.77& 25.40 25.27& 25.90 25.90& 26.40 26.22& 26.90 -----& 27.40 -----& 27.90 -----& 28.40 -----& 28.90 -----\\	     
26  &25.07 24.91& 25.57 25.46& 26.07 25.95& 26.57 26.29& 27.07 -----& 27.57 -----& 28.07 -----& 28.57 -----& 29.07 -----\\	     
28  &25.23 25.08& 25.73 25.57& 26.23 26.21& 26.73 26.59& 27.23 -----& 27.73 -----& 28.23 -----& 28.73 -----& 29.23 -----\\	     
30  &25.38 25.20& 25.88 25.74& 26.38 26.20& 26.88 -----& 27.38 -----& 27.88 -----& 28.38 -----& 28.88 -----& 29.38 -----\\	     
32  &25.52 25.30& 26.02 25.78& 26.52 26.32& 27.02 -----& 27.52 -----&
28.02 -----& 28.52 -----& 29.02 -----& 29.52 -----\\ 
\hline
\end{tabular}
\end{table*}

\label{lastpage}

\end{document}